\documentclass[%
 reprint,
tightenlines,
showpacs,preprintnumbers,
 amsmath,amssymb,
 aps,
 pra,
]{revtex4-1}

\usepackage{graphicx}
\usepackage{dcolumn}
\usepackage{bm}
\usepackage{hyperref}
\usepackage{breakurl}

\usepackage{CJKutf8}
\newenvironment{Chinese}{%
\CJKfamily{bsmi}%
\CJKtilde
\CJKnospace}{} 

\renewcommand{\ni}{n_{\rm ini}}

\begin{document}

\preprint{APS/123-QED}

\title{Spinor Bose-Einstein Condensates of Positronium}

\author{Yi-Hsieh Wang\begin{CJK}{UTF8}{}\begin{Chinese} (汪以謝)\end{Chinese}\end{CJK}}
\author{Brandon M. Anderson\begin{CJK}{UTF8}{}\begin{Chinese} (安博仁)\end{Chinese}\end{CJK}}
\author{Charles W. Clark\begin{CJK}{UTF8}{}\begin{Chinese} (查理斯 克拉恪)\end{Chinese}\end{CJK}}
\affiliation{%
 Joint Quantum Institute, University of Maryland College Park \\ 
and National Institute of Standards and Technology. 
}%

\begin{abstract}
Bose-Einstein condensates (BECs) of positronium (Ps) have been of experimental and theoretical interest due to their potential application as the gain medium of a $\gamma$-ray laser. Ps BECs are intrinsically spinor due to the presence of ortho-positronium (\textit{o}-Ps) and para-positronium (\textit{p}-Ps), whose annihilation lifetimes differ by three orders of magnitude. In this paper, we study the spinor dynamics and annihilation processes in the \textit{p}-Ps/\textit{o}-Ps system using both solutions of the time-dependent Gross-Pitaevskii equations and a semiclassical rate-equation approach. The spinor interactions have an $O(4)$ symmetry which is broken to $SO(3)$ by an internal energy difference between \textit{o}-Ps and \textit{p}-Ps. For an initially unpolarized condensate, there is a threshold density of $\approx 10^{19}$ cm$^{-3}$ at which spin mixing between \textit{o}-Ps and \textit{p}-Ps occurs. Beyond this threshold, there are unstable spatial modes accompanied by spin mixing. To ensure a high production yield above the critical density, a careful choice of external field must be made to avoid the spin mixing instability.
\end{abstract}


\maketitle

\section{\label{sec:level1}Introduction\\}

Positronium (Ps) is an atom composed of one electron and one positron.  Its energy levels are similar to those of atomic hydrogen on a reduced energy scale, and the total spin angular momentum, $S = 0$ or 1, is approximately conserved.  The Ps ground state has no orbital angular momentum, and has $^1S$ (para-positronium, para- or \textit{p}-Ps) and $^3S$ (ortho-positronium, ortho- or \textit{o}-Ps) components that are separated by an energy $\epsilon=1.351\times 10^{-22}\; {\rm J}$ \cite{Rich, Charlton}.  The electron and positron annihilate each other and emit $\gamma$ rays. The \textit{p}-Ps state can decay by emitting two $\gamma$ rays, and has a lifetime  of $\tau_{\rm{p}} = 0.125\; \mathrm{ns}$; the \textit{o}-Ps state must emit at least three $\gamma$ rays, and has a lifetime of $\tau_{\rm{o}} = 0.142~ \mu\mathrm{s}$.

In 1994 Platzman and Mills \cite{MillsJr1994} considered the possibility of using \textit{o}-Ps atoms to make a Ps Bose-Einstein condensate (BEC), which has potential application as the gain medium of a $\gamma$-ray laser \cite{MillsJr2002107,Mills2004}. In their proposal, Ps atoms can be created and collected by impinging a dense positron pulse into a cavity within a piece of silicon.  A fraction of the positrons will capture an electron in the silicon to yield Ps. It is estimated that at a temperature of 15 K, a BEC of \textit{o}-Ps will be formed if the Ps density exceeds $10^{18}$ cm$^{-3}$ \cite{Mills2004}. Moreover, by converting the long-lived \textit{o}-Ps condensate to a \textit{p}-Ps condensate with a magnetic field, strong $\gamma$-ray emission can be generated as the outcome of annihilation of \textit{p}-Ps. At present, it is possible to generate a spin-polarized Ps gas with phase-space density about two orders of magnitude smaller than the condensation requirement \cite{Cassidy2010}. Further improvements in the trapping, cooling, and polarizing techniques \cite{Cassidy2008, Cassidy2010, Cassidy2007} may make possible the formation of a Ps BEC in the future.

There are some differences between Ps BEC and the BECs of alkali atoms. The mass of a Ps atom ($m=2m_\mathrm{e}$) is much less than the mass of ordinary atoms, and thus Ps can be brought to Bose-Einstein condensation at much higher number densities, $n$, and/or temperatures, $T$, than are required for laser cooled atoms \cite{Pethick}. Ps also has a small scattering length $a\sim 0.1 \textrm{ nm}$, so that a Ps condensate remains a weakly interacting dilute Bose gas in the higher density regime, i.e., ($n a^{3} \ll 1$). Second, a Ps BEC has a short metastable lifetime $\tau \sim \tau_{\rm o}$ due to spontaneous annihilation into $\gamma$ rays. Despite its short life, a Ps BEC can be observed though gamma-ray emission signatures \cite{MillsJr2002107}. As will be discussed later, an appropriately prepared condensate can be used as the gain medium for  $\gamma$ ray laser, which can be observed through the coherent emission of $\gamma$ rays.

In addition to the $\gamma$-ray laser, Ps BECs can also be used to make a Ps atom laser. It has been proposed that for a cavity filled with Ps BEC, Ps atoms can tunnel through the small holes in the cavity wall and propagate into vacuum as a monoenergetic Ps beam \cite{Cassidy2007, MillsJr2007}. A Ps atom laser has applications to fundamental studies of antimatter. For example, it could be applied in high-precision measurements of the Ps $1^3S_1-2^3S_1$ transition \cite{MillsJr2007}. A Ps atom laser could also be applied to gravity-related interference experiments. A Mach-Zender interferometer based on beams consisting of high-Rydberg Ps atoms can be set up in a vertical plane to measure the gravity-dependent phase shift \cite{Cassidy2007}. Although it is of interest to study the implementation of an atom laser using Ps  BEC, in this paper we focus on the spinor properties of Ps BEC and its application to the $\gamma$-ray laser.

In the preparation of dense Ps, it has been shown \cite{Cassidy2005,Cassidy2010} that ortho-to-para spin exchange is one of the main mechanisms for quenching the \textit{o}-Ps population. To avoid this population loss, a polarized positron beam should be used to generate a nonzero fraction of polarized \textit{o}-Ps that is long-lived. During the condensation process spin-mixing collisions convert the unpolarized fraction of \textit{o}-Ps to \textit{p}-Ps, which quickly annihilates, and the remaining \textit{o}-Ps is polarized. A standard design of a $\gamma$-ray laser based on a Ps BEC uses polarized \textit{o}-Ps as a storage medium, which is quickly converted to \textit{p}-Ps by a magnetic field switch that triggers stimulated annihilation \cite{Mills2004}. After applying the field, the system again obtains an unpolarized fraction that may undergo spin mixing and reduce the final $\gamma$-ray yield. In this paper, we study the time evolution of \textit{o}-Ps and \textit{p}-Ps mixtures in order to understand the interplay of spin-mixing collision rates and $\gamma$-annihilation rates. We study spin mixing and Ps self-annihilation effects using two approaches: the solution of the time-dependent Gross-Pitaevskii (GP) equations that described a mixture of o- and \textit{p}-Ps BECs, and a semiclassical rate-equation method.  We also consider how to avoid para- to ortho- conversion during laser action. 

To understand the physical properties of a Ps BEC, especially the interplay of spin mixing and Ps annihilation, we start with the GP formalism that consists of both para- and ortho- states. The formalism has the structure of a mixture of spin-1 and spin-0 BECs, and the full spin-mixing interactions have an $O(4)$ symmetry \cite{Judd1975}. The internal energy splitting breaks this symmetry to a $SO(3)$ symmetry among the triplet states. There is a competition between the internal energy splitting and spin-mixing terms in the second-quantized Hamiltonian, which determines the ground state phase, the spin-mixing dynamics, and the stability of the system. At low density, the ground state phase consists of pure \textit{p}-Ps. There is a critical density above which spin mixing becomes significant and the ground state acquires a non-zero \textit{o}-Ps fraction. We believe our present work is the first to have investigated the spinor structure of Ps BECs. This structure may also be relevant to BECs of hydrogen or tritium \cite{Willian1976, Blume2002}. 

Our paper is organized as follows. Sec. \ref{sec:GP} starts with a formalism based on the Ps second-quantized Hamiltonian which is used to derive the time-dependent GP equations. In Sec. \ref{sec:sym} we study the symmetry properties of the Ps system that are invariant under spin rotation. In Sec. \ref{sec:stat}, we show that there is a phase transition in the ground state composition at a critical density $n_{\rm c}$. Spin mixing and dynamic instability effects become significant for densities greater than $n_{\rm c}$. In Sec. \ref{sec:spin-mix}, we study the spin-mixing dynamics of a mixed homogeneous BEC numerically and analytically at densities above and below $n_{\rm{c}}$. Sec. \ref{sec:annihilation} considers the annihilation of Ps under the influence of spin mixing. We propose a scheme that optimizes the $\gamma$-ray laser from a polarized \textit{o}-Ps BEC. The rate equation approach is first used to model the dynamics for an incoherent mixture. Then, we apply the time-dependent GP equations to simulate the time evolutions of a Ps condensate with  decay. At the same time, dynamic stability is taken into account through the Bogoliubov equations and tested numerically through the GP equations with spatial random noises. The last section summarizes our results.

\section{\label{sec:GP} Gross-Pitaevskii Theory for Positronium condensates}

There are many previous studies of the interactions of two Ps atoms, including their fusion into the diatomic Ps molecule \cite{Schrader2004,Ivanov2001,Ivanov2002}.  We use results of some of this previous work to determine the scattering lengths for low-energy Ps collisions that are relevant to describing a BEC of Ps within the conventional mean-field theoretical framework: the time-dependent Gross-Pitaevskii (GP) equation.
In particular, we use the Gross-Pitaevskii theory to study physical properties of a mixture of \textit{o}-Ps and \textit{p}-Ps condensates, such as the symmetry under spin rotation operations, the stationary structure, and the dynamics. The treatment here ignores the effects of electron-positron annihilation. These are included in subsequent sections.

There are four spin states of Ps which we designate by $|{\rm{p}}\rangle$, $|1\rangle$, $|0\rangle$, $|-1\rangle$, where  $|{\rm{p}}\rangle$ is the \textit{p}-Ps state, and $|M\rangle$ is the \textit{o}-Ps state with spin projection of $M\hbar$ upon the $\hat{z}$-axis for $M=1,0,-1$. As shown in Appendix \ref{app:Hint}, the second-quantized interaction Hamiltonian for this system takes the following form:
\begin{eqnarray}
\label{eq:H_int}
\mathcal{H}_{\rm{int}}&&=
\frac{g_{0}}{2} \sum_{\substack{i,j=\\0, \pm 1, {\rm{p}}}} \int d^3 { r} \,
\Psi_{i}^{\dagger} \Psi_{j}^{\dagger} \Psi_{i} \Psi_{j} \nonumber\\
&&+\frac{g_{1}}{2} \int d^3 {r} \,
\left( \begin{array}{c}
  2\Psi_{1}\Psi_{-1}\\ 
  -\Psi_{0}\Psi_{0}\\
  \Psi_{{\rm{p}}}\Psi_{{\rm{p}}}
\end{array}\right)^{\dagger}
\left( \begin{array}{ccc}
  1 &  1 & 1 \\ 
  1 &  1 & 1 \\ 
  1 &  1 & 1 \\ 
\end{array}\right)
\left( \begin{array}{c}
  2\Psi_{1}\Psi_{-1}\\ 
  -\Psi_{0}\Psi_{0}\\
  \Psi_{{\rm{p}}}\Psi_{{\rm{p}}}
\end{array}\right),\nonumber\\ 
\end{eqnarray} 
where $\Psi_{i}(\mathbf{r}) \left(\Psi_{i}^\dagger(\mathbf{r})\right)$ annihilates(creates) a particle of spin state $|i\rangle$ at point $\mathbf{r}$, and $g_{0}$ and $g_1$ are pseudo-potential constants with values $g_0/\hbar=1.154\times 10^{-7}\;\rm{cm}^3/\rm{s}$ and $g_{1}/\hbar= 5.240\times 10^{-8}\;\rm{cm}^3/\rm{s}$. The full many-body Hamiltonian, including the single-particle contribution, is 
\begin{eqnarray}
\label{eq:H}
\mathcal{H}&&=\int d^3 { r} \sum\limits_{i=0,\pm1, {\rm{p}}} 
\Psi_{i}^{\dagger}\left(\frac{\mathbf{p}^2}{2m}+V_{\rm{ext}}+\epsilon_i\right)\Psi_{i}+\mathcal{H}_{\rm{int}} \nonumber\\
\end{eqnarray} 
where $\mathbf{p}$ is the momentum operator, $m$ is the mass of Ps, $V_{\rm{ext}}$ is the external potential, and $\epsilon_i$ is the internal energy of spin state $i$. Throughout this paper, $\epsilon_1=\epsilon_0=\epsilon_{-1}\equiv\epsilon_{\rm{o}}$. 

Now we go to the mean field limit by replacing each field operator $\Psi_{i}$ by the corresponding mean field value $\psi_{i}=\langle \Psi_i \rangle$ \cite{Pethick}. The interaction term can be expressed as 
\begin{eqnarray}
\label{eq:H_int2}
\mathcal{H}_{\rm{int}}=\frac{1}{2}\int d^3 {r} \,\left(g_0 n^{2}+g_1 \left|2\psi_1 \psi_{-1} -\psi_0^{2}+ \psi_{\rm{p}}^{2}\right|^{2}\right)
\end{eqnarray}
where $n=\sum_{i=0,\pm1,{\rm{p}}} |\psi_i|^2$ denotes the total number density. The equations of motion can be found by taking the functional derivative of the mean-field Hamiltonian 
\begin{eqnarray}
\label{eq:eom}
i\,\hbar \dot{\psi}_{i}=\frac{\delta\mathcal{H} }{\delta \psi_i^*}.
\end{eqnarray}
This yields the time-dependent GP equations:
\begin{eqnarray}
\label{eq:GP}
i\,\hbar \dot{\psi}_{1} &=& \left(H_0+\epsilon_{\rm{o}}+2g_1|\psi_{-1}|^{2} \right) \psi_1+g_1 \psi_{-1}^{*}(\psi_{{\rm{p}}}^{2}-\psi_{0}^{2})\nonumber\\
i\,\hbar \dot{\psi}_{0} &=& \left(H_0+\epsilon_{\rm{o}}+g_1|\psi_{0}|^{2} \right) \psi_0-g_1\psi_{0}^{*}(2\psi_{1}\psi_{-1}+\psi_{{\rm{p}}}^{2})\nonumber\\
i\,\hbar \dot{\psi}_{-1} &=& \left(H_0+\epsilon_{\rm{o}}+2g_1|\psi_{1}|^{2} \right) \psi_{-1}+ g_1\psi_{1}^{*}(\psi_{{\rm{p}}}^{2}- \psi_{0}^{2})\nonumber\\
i\,\hbar \dot{\psi}_{{\rm{p}}} &=& \left(H_0+\epsilon_{\rm{p}}+g_1|\psi_{{\rm{p}}}|^{2} \right) \psi_{\rm{p}}+g_1\psi_{{\rm{p}}}^{*}(2\psi_{1}\psi_{-1}-\psi_{0}^{2})\nonumber\\
\end{eqnarray}
where
\begin{eqnarray}
\label{H_i}
H_0= \frac{\mathbf{p}^2}{2m}+V_{\rm{ext}}+g_{0}n.
\end{eqnarray}
The $g_1$ terms in the second half of the RHS of Eq. \ref{eq:GP} are responsible for the population exchange among spin states. For the remainder of this section, we take $\epsilon_{\rm{o}}=\epsilon$ and $\epsilon_{\rm{p}}=0$, which amounts to ignoring the spontaneous annihilation process.  

\subsection{\label{sec:sym}Symmetry under spin rotations}
We now consider the symmetry of the system including interactions. We can express $\mathcal{H}_{\rm{int}}$ in terms of the ortho--para- spinor $\psi^T=\left(\psi_1,\;\psi_0,\;\psi_{-1},\;\psi_{\rm{p}}\right)$ 
\begin{eqnarray}
\label{eq:Hq}
\mathcal{H}_{\rm{int}}=\frac{1}{2} \int d {r^3} \,\left( g_0 n^{2}+ g_1\left|\psi^{T}Q\psi\right|^2\right),
\end{eqnarray}
where
\begin{eqnarray}
\label{eq:M}
Q=\left(\begin{array}{cccc}
0 & 0 & 1 & 0\\
0 & -1 & 0 & 0\\
1 & 0 & 0 & 0\\
0 & 0 & 0 & 1
\end{array}\right).
\end{eqnarray}
In this basis, the components of the spin operator $ \mathbf{S}  = S_a \hat{\mathbf{e}}_a$, with $a=1,2,3$, are expressed as 
\begin{eqnarray}
S_{1} & = & \frac{1}{\sqrt{2}}\left(\begin{array}{cccc}
0 & 1 & 0 & 0\\
1 & 0 & 1 & 0\\
0 & 1 & 0 & 0\\
0 & 0 & 0 & 0
\end{array}\right),\\
S_{2} & = & \frac{1}{\sqrt{2}}\left(\begin{array}{cccc}
0 & -i & 0 & 0\\
i & 0 & -i & 0\\
0 & i & 0 & 0\\
0 & 0 & 0 & 0
\end{array}\right),\\
S_{3} & = & \left(\begin{array}{cccc}
1 & 0 & 0 & 0\\
0 & 0 & 0 & 0\\
0 & 0 & -1 & 0\\
0 & 0 & 0 & 0
\end{array}\right).
\end{eqnarray}
It is straightforward to show that $[S_a,Q]=0$ for all $a$.

In the limit that $\psi_{\rm{p}} = 0$, the interaction Hamiltonian is identical to that for a normal spin-1 spinor condensate~\cite{Ho1998, Machida}. That is
\begin{eqnarray}
\label{eq:H_int3}
\mathcal{H}_{\rm{int}}&=&\frac{1}{2}\int d^3 { r} \,\left(g_{0} n^{2}+g_1 n^2\left( 1-\langle \bold{S} \rangle \cdot \langle \bold{S} \rangle \right)\right)
\end{eqnarray}
where $ \left\langle S_a \right\rangle = \psi^\dagger S_{a} \psi/n$ with $a=1,2,3$ is the average spin per atom, so that $0\leq\langle \mathbf{S} \rangle \cdot \langle \mathbf{S} \rangle\leq 1$. Since $g_1>0$, the ground state will have $\langle \mathbf{S} \rangle \cdot \langle \mathbf{S} \rangle=1$ in the limit $\psi_{\rm{p}}=0$. Thus, in this limit, the ortho- sector is a ferromagnetic condensate. However, when $\psi_{\rm{p}}\neq0$, the ground state has $\langle \mathbf{S} \rangle =0$, as will be shown in the next section. As we shall see, in parameter regimes of current experimental interest, Eq. (\ref{eq:Hq}) can induce significant interconversion between \textit{o}-Ps and \textit{p}-Ps.

To specify the symmetry of the system, we can use $\mathbf{S}$ as the generator of rotations $D_\mathbf{\hat{n}}(\alpha) = e^{-i\alpha \hat{\mathbf{n}} \cdot \mathbf{S}}$ among the spin-1 states, where $\alpha$, $\mathbf{\hat{n}}$ denote the angle and axis of the rotation. Since $\psi^T Q \psi$ is invariant under $\psi\rightarrow e^{i\alpha S_{a}}\psi$, we see that $\mathcal{H}_{\rm{int}}$ is also invariant under arbitrary spin rotations in the ortho- sector, and  $\langle S_a \rangle$ for each $a$ is a conserved quantity of the system. This implies that from a given solution to Eq. (\ref{eq:GP}), we can obtain a manifold of equivalent solutions related by rotations of the form $D_{\hat{\bf{n}}}(\alpha)$. In particular, this implies that there is a continuous degeneracy of the many-body ground state.   

Rotations between the ortho- sector and the para- sector $e^{-i\alpha \hat{\mathbf{n}} \cdot \mathbf{R}}$ can be generated by operators $ \mathbf{R}  = R_a \hat{\mathbf{e}}_a$ with $a=1,2,3$, and
\begin{eqnarray}
R_{1} & = & \frac{1}{\sqrt{2}}\left(\begin{array}{cccc}
0 & 0 & 0 & 1\\
0 & 0 & 0 & 0\\
0 & 0 & 0 & -1\\
1 & 0 & -1 & 0
\end{array}\right),\\
R_{2} & = & \frac{1}{\sqrt{2}}\left(\begin{array}{cccc}
0 & 0 & 0 & -i\\
0 & 0 & 0 & 0\\
0 & 0 & 0 & -i\\
i & 0 & i & 0
\end{array}\right),\\
R_{3} & = & \left(\begin{array}{cccc}
0 & 0 & 0 & 0\\
0 & 0 & 0 & -1\\
0 & 0 & 0 & 0\\
0 & -1 & 0 & 0
\end{array}\right).
\end{eqnarray}
It can be shown that $\mathcal{H}_{\rm{int}}$ is also invariant under rotations $\psi\rightarrow e^{i\alpha R_{a}}\psi$. As shown in Appendix B, the commutation relations among $S_a$ and $R_b$ are those of the $O(4)$ rotation group. This symmetry is broken in the full Hamiltonian due to the one-body internal energy difference between ortho- and para- Ps. The only global symmetry is that generated by the operators  $S_a$ \cite{Ho1998}.

\subsection{\label{sec:stat}Ground State of a Ps BEC}
We now consider the ground state of a Ps BEC in the zero temperature limit. Suppose there exists a stationary state $\psi_i(\mathbf{r},t)=\sqrt{n_{i}(\mathbf{r})}e^{i\phi_{i}(\mathbf{r})}e^{-i\mu t/\hbar}$, where $n_i$ and $\phi_i$ are the number density and phase for the component $i$, and $\mu$ is the chemical potential. Substituting this ansatz into Eq. \ref{eq:GP}, we can calculate the equilibrium composition of the ground state.
In proposed implementations of Ps BEC,~\cite{MillsJr1994,MillsJr2002107} Ps is collected in a cavity of volume $\sim 10^{-13}\textrm{ cm}^{3}$ with a range of densities between $10^{18}\textrm{ cm}^{-3}$ and $10^{21}\textrm{ cm}^{-3}$. Given the minimal required densities for condensation, the largest attainable values of the healing length, $\xi = \hbar /\sqrt{2 m \mu}$, are about $\xi \sim 10^{-6} \textrm{ cm}$. This is much smaller than the characteristic cavity dimension $\sim 10^{-4}\textrm{ cm}$. For this reason, we will neglect the contribution of the kinetic energy operator $\mathbf{p}^2/2m$ for the remainder of this section so that the condensate is uniform in space.

To understand the ortho-/para- balance of the condensate, we consider the competition between the $g_1$ interaction and the energy separation $\epsilon$, which is manifested in the energy as
\begin{eqnarray}
\label{eq:E1}
E &=& \int d^3 {r} \,\left(\epsilon n_{\rm{o}}+\frac{g_1}{2}\left|2\psi_1 \psi_{-1} -\psi_0^{2}+ \psi_{\rm{p}}^{2}\right|^{2}\right),
\end{eqnarray}
where $n_{\rm{o}}=n_1+n_0+n_{-1}$. We rewrite the term that is quartic in $\psi_i$ as
\begin{eqnarray}
\label{eq:E2}
\frac{g_1}{2}\left|2\sqrt{n_1 n_{-1}}e^{i(\phi_1+\phi_{-1})}-n_0 e^{2i\phi_0}+ n_{\rm{p}} e^{2i\phi_{\rm{p}}}\right|^{2}.
\end{eqnarray}

\begin{figure}
 \includegraphics[width=3.3in]{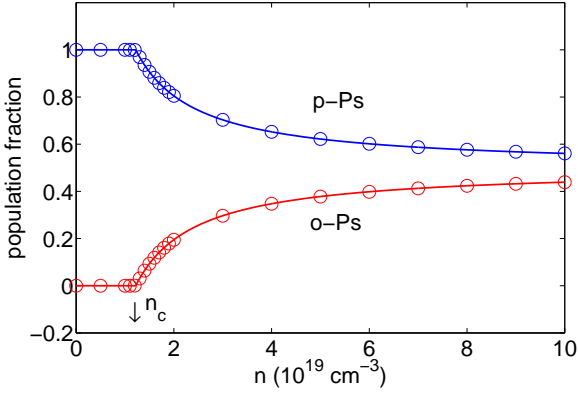}
\caption{\label{fig:stationary} Relative population fraction of ortho-/para- sectors as a function of density. The solid lines correspond to Eq. \ref{eq:singlet}, and the circled dots are obtained by using an imaginary-time approach for the GP equation. The critical density $n_{\mathrm{c}}\approx 1.2 \times 10^{19}$cm$^{-3}$ corresponds to a nonzero occupation of the ortho- sector in the ground state. Below the critical density, the ground state is a pure \textit{p}-Ps condensate. In the high density limit, $n\gg n_{\mathrm{c}}$, the ratio of $n_{\rm{p}}/n_{\rm{o}} \rightarrow 1$.}
\end{figure}

\begin{figure}
 \includegraphics[width=3.3in]{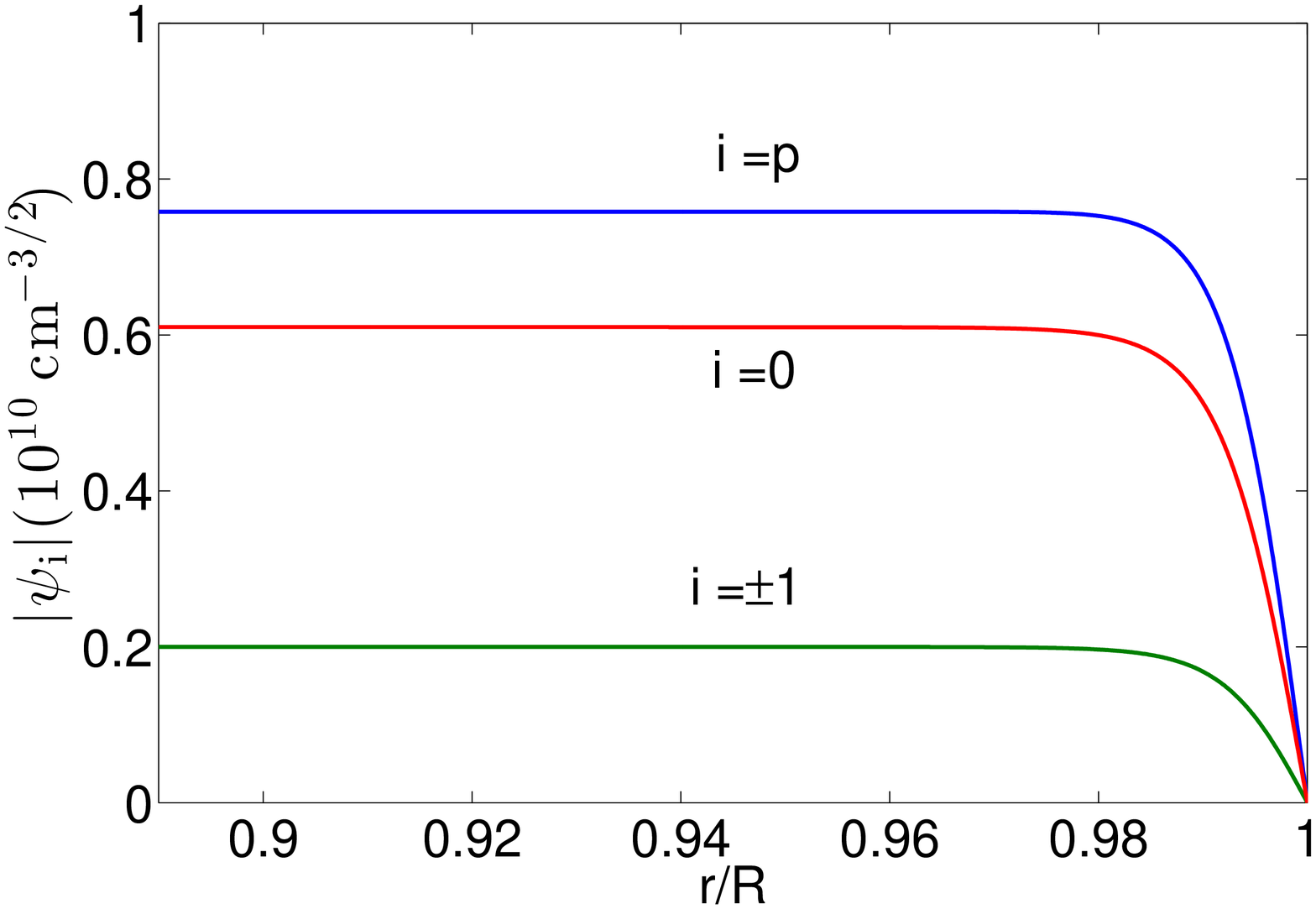}
\caption{\label{fig:stationary2} $\left|\psi_i(r)\right|$ for the ground state of Ps BEC at $n=10^{20}$ cm$^{-3}$ confined in a spherical cavity of volume $4\pi R^3/3=10^{-13}$ cm$^{3}$. These results, obtained from integration of Eq. \ref{eq:GP} in imaginary time, demonstrate the uniformity and miscibility of the mixed condensates and the equality $\left|\psi_1\right|=\left|\psi_{-1}\right|$. The density variation is confined to a boundary layer with thickness about $1\%$ of the cavity radius $R$. Starting from random initial conditions, the numerical calculation converges on a solution that has the phase relationship indicated in Eq. \ref{eq:psi3}. }
\end{figure}

The effect of $\epsilon>0$ is to suppress $n_{\rm{o}}$ in the ground state. The quartic term on the other hand is minimized at nonzero $n_{\rm{o}}$. In the limit of $\epsilon \gg g_1 n$, we obtain a pure \textit{p}-Ps BEC, $n_{\rm{o}}\rightarrow 0$. In the other limit, $\epsilon \ll g_1 n$, the ground state energy is dominated by the quartic interaction, which vanishes when
\begin{eqnarray}
\label{eq:psi}
n_{\rm{p}} e^{2i\phi_{\rm{p}}}=n_0 e^{2i\phi_0}-2\sqrt{n_1 n_{-1}}e^{i(\phi_1+\phi_{-1})}.
\end{eqnarray}
Since $\epsilon_{\rm{p}}<\epsilon_{\rm{o}}$, the ground state has the largest value of $n_{\rm{p}}$ consistent with Eq. \ref{eq:psi} and the fixed value of $n=n_{\rm{o}}+n_{\rm{p}}$. It can be seen from Eq. \ref{eq:psi} that $n_{\rm{p}}$ is maximized when $\phi_{\rm{p}}+q\pi=\phi_0=[\phi_1+\phi_{-1}+(2r+1)\pi]/2$ for integers $q,r$, and thus $n_{\rm{p}}=n_0+2\sqrt{n_1 n_{-1}}$. To see if there exists a number imbalance between $n_1$ and $n_{-1}$, we let $n_1=\bar{n}+m$ and $n_{-1}=\bar{n}-m$, where $\bar{n}$ represents the average density of the two species, and $m$ denotes the number imbalance. Keeping $n_1+n_{-1}$ a constant while varying $m$, we find that $\sqrt{n_1 n_{-1}}=\sqrt{\bar{n}^2-m^2}$ has the largest value if the imbalance $m=0$. Therefore, we obtain the maximal \textit{p}-Ps density 
\begin{eqnarray}
\label{eq:psi2}
n_{\rm{p}} =n_0+2\bar{n}=n_{\rm{o}},
\end{eqnarray}
so ortho- and para- populations become equal in the high-density limit. The ortho- sector is equivalent to the polar state of a spin-1 condensate with $ \left\langle \mathbf{S} \right\rangle = 0$ ~\cite{Ho1998,Masahito}, for which there exists a degree of freedom to distribute the population between $n_0$ and $\bar{n}$. Adapting the parametrization given by Ho \cite{Ho1998}, which has since become standard \cite{Masahito,Kawaguchi2012253}, we find the general expression for the ground state to be, up to an overall phase,
\begin{eqnarray}
\label{eq:psi3}
\psi & = & \sqrt{n_{\rm{o}}}\left(\begin{array}{c}
-\frac{1}{\sqrt{2}}e^{-i\alpha}\cos(\beta)\\
\sin(\beta)\\
\frac{1}{\sqrt{2}}e^{i\alpha}\cos(\beta)\\
0\end{array}\right)+ \sqrt{n_{\rm{p}}}\left(\begin{array}{c}
0\\0\\0\\ \pm 1
\end{array}\right)\nonumber\\
\end{eqnarray}
where $\alpha$, $\beta$ are arbitrary real numbers. To fully specify the ground state of the system, we now identify the relationship between $n_{\rm{o}}$ and $n_{\rm{p}}$. 

When the effects associated with $\epsilon$ and the quartic term in Eq. \ref{eq:E1} are comparable,  $n_{\rm{p}}$ will lie in the range $n/2<n_{\rm{p}}<n$. To determine $n_{\rm{p}}$, we write $n_{\rm{p}}=n_{\rm{o}}+\delta n$ and adopt the phases and densities described in our derivation of Eq. \ref{eq:psi3}, then Eq. \ref{eq:E1} takes the form 
\begin{eqnarray}
\label{eq:E3}
E = \int d^3 {r} \,\left( \epsilon n_{\rm{o}}+\frac{g_1}{2} \delta n^2  \right).
\end{eqnarray}
Using $n_{\rm{o}}=(n-\delta n)/2$, we find $n_{\rm{p}}$ by minimizing Eq. \ref{eq:E3} with respect to $\delta n$, for fixed $n$. We obtain
\begin{eqnarray}
\label{eq:singlet}
\begin{array}{ll}
 n_{\rm{p}}/n=  \frac{1}{2}(1+ \frac{n_{\mathrm{c}}}{n} ) & \mathrm{if} \;\;  n> n_{\mathrm{c}}\\
 n_{\rm{p}}/n= 1 & \mathrm{if} \;\;  n \leq n_{\mathrm{c}}
\end{array} 
\end{eqnarray}
where the critical density,
\begin{eqnarray}
\label{eq:nc}
 n_{\mathrm{c}}= \frac{\epsilon}{2 g_1}, 
\end{eqnarray}
is that at which the para- fraction starts to depart from 1. For Ps, $\epsilon$ is known experimentally and theoretically \cite{Rich} and $g_1$ has been determined through first-principles calculations, as summarized in Appendix \ref{app:Hint}. With these values, we find $n_{\mathrm{c}}\approx  1.2 \times 10^{19}$ cm$^{-3}$. 

We have verified this simple model by exact numerical calculation of the ground state of a Ps BEC as a function of density by integration of the GP equations, Eq. \ref{eq:GP}, in imaginary time. As shown in Fig. \ref{fig:stationary}, the results are consistent with Eq. \ref{eq:singlet}.

The results of this section are based on the premise that individual condensates are uniform and miscible throughout the cavity. To verify the validity of this picture for Ps confined in a cavity, we impose a hard-wall boundary condition on Eq. \ref{eq:GP} and calculate the ground state by integrating in imaginary time. For an isotropic cavity of volume $10^{-13}$ cm$^{3}$ and total density $n=10^{20}$ cm$^{-3}$, we obtain the groud-state solution of a mixed condensate given in Fig. \ref{fig:stationary2}. We see that the assumption holds nicely in the figure with very small boundary effect. The individual condensates are miscible and uniform throughout the bulk region, and the spatial variation around the boundary is only about 0.01 of the cavity radius. In addition, the stationary populations obtained here coincide with those shown in Fig. \ref{fig:stationary}. 

\subsection{\label{sec:spin-mix}Spin-mixing dynamics}

If a Ps BEC is prepared in a non-stationary state, the interaction Hamiltonian $\mathcal{H}_{\rm{int}}$ can lead to spin-mixing dynamics \cite{Pu1999, Masahito, Kawaguchi2012253}. To understand the basics of spin-mixing dynamics and its dependence on the critical density $n_{\mathrm{c}}$, we neglect the contribution of the kinetic energy operator and treat the condensate as uniform. GP calculations of nonuniform condensates are included in the next section, where we find pronounced effects of inhomogeneity for $n>n_{\mathrm{c}}$.

It is convenient to reformulate the coupled GP equations in terms of the fractional populations of individual spin states and the relative phases. For the  simplified case involving only the wavefunctions $\psi_{\rm{p}}=\sqrt{n\rho}e^{i\phi_{\rm{p}}}$ and $\psi_0 = \sqrt{n(1-\rho)}e^{i\phi_0}$, the equations of motion can be recast as
\begin{eqnarray}
\label{eq:Josephson}
\frac{d \rho}{d \tau}&=&\nu\rho(1-\rho)\sin 2\phi,\nonumber\\
\frac{d \phi}{d \tau}&=&\frac{\nu}{2}(1-2\rho)(1+\cos 2\phi )+1,
\end{eqnarray}
where $\phi = \phi_{\rm{p}}-\phi_{0}$ is the relative phase between the two components, $\rho$ is the population fraction in the para- component, $\tau=\epsilon t/\hbar$, and $\nu=n/n_{\mathrm{c}}$ (from Eq. \ref{eq:nc}). As in the Josephson effect~\cite{Raghavan1999}, the phase and density differences drive population oscillations between the two species. Using a method from Ref. \cite{Zhang2005}, Eq. \ref{eq:Josephson} can be solved by recognizing that $\rho$ and $\phi$ are conjugate variables of a functional,
\begin{eqnarray}
\label{eq:E'}
\mathcal{E}= \frac{\nu}{2}\rho(1-\rho)(1+\cos 2\phi) + \rho,
\end{eqnarray}
which is a constant of motion determined by the initial values $\rho(\tau_0)$ and $\phi(\tau_0)$ at an arbitrary value of the scaled time $\tau=\tau_0$. Combining Eq. \ref{eq:E'} and Eq. \ref{eq:Josephson}, we obtain
\begin{eqnarray}
\label{eq:GP2}
\left(\frac{d \rho}{d \tau}\right)^2&=& 4 \nu \left(\rho-\mathcal{E}\right)(\rho-\rho_+)(\rho-\rho_-)
\end{eqnarray}
where
\begin{eqnarray}
\label{eq:GP2sol}
\rho_\pm = \frac{1}{2\nu}\left(1+\nu\pm \sqrt{\left(1+\nu\right)^2-4\mathcal{E}\nu}\right).
\end{eqnarray}
As shown in \cite{Zhang2005, Hancock}, this differential equation can be solved by the Jacobi elliptic~\cite{NIST:DLMF,Olver:2010:NHMF} function sn (or cn) as 
\begin{eqnarray}
\label{eq:GP2sol2}
\rho = \rho_3+(\rho_2-\rho_3) \mathrm{sn}^2 \left(\sqrt{\nu(\rho_1-\rho_3)}\; \tau, k_1 \right)\nonumber\\
\end{eqnarray}
where $k_1$  is the elliptic modulus given by
\begin{eqnarray}
k_1^2=\frac{\rho_2-\rho_3}{\rho_1-\rho_3}
\end{eqnarray}
and $\rho_i$ are the three zeros of Eq. \ref{eq:GP2}, ($\rho_i=\mathcal{E},\rho_{\pm}$), ordered so that $\rho_1>\rho_2>\rho_3$. Eq. \ref{eq:GP2sol2} is strictly periodic in variable $\tau$, in the limit $k_1\rightarrow0$ it becomes sinusoidal. The period of the solution, $\rm{T}$, can be calculated in terms of the elliptic integral of the first kind $F(\phi, k)$ \cite{NIST:DLMF,Olver:2010:NHMF} as
\begin{eqnarray}
{\rm{T}}=\frac{2}{\sqrt{\nu\left(\rho_1-\rho_3\right)}}F\left(\frac{\pi}{2}, k_1\right).
\end{eqnarray}
We have chosen the origin $\tau=0$, so that $\rho(0)=\rho_3$, the minimum value of para- population fraction. The population oscillates between $\rho_3$ and $\rho_2$. 
\begin{figure}
 \includegraphics[width=3.3in]{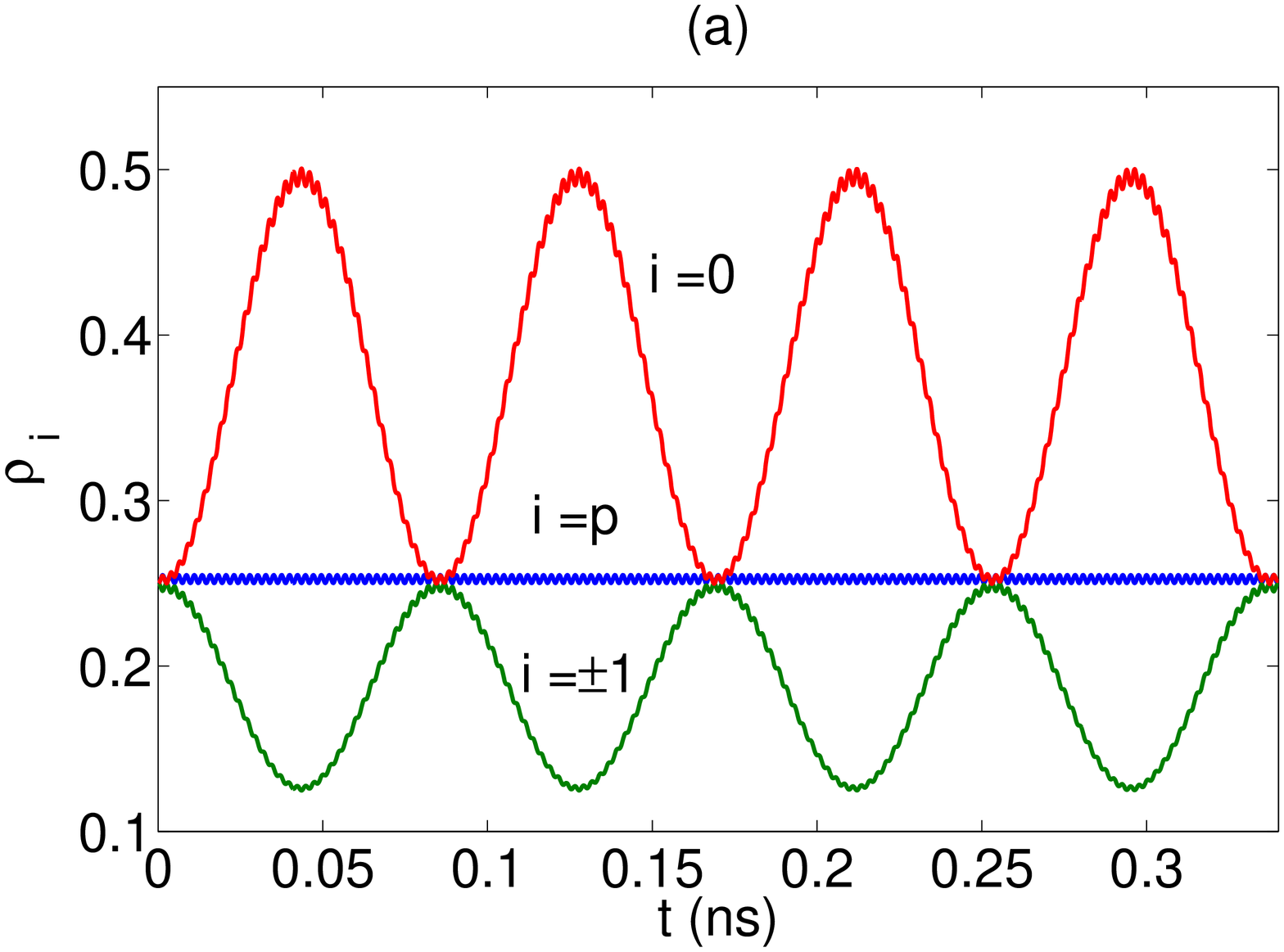}\\
 \includegraphics[width=3.3in]{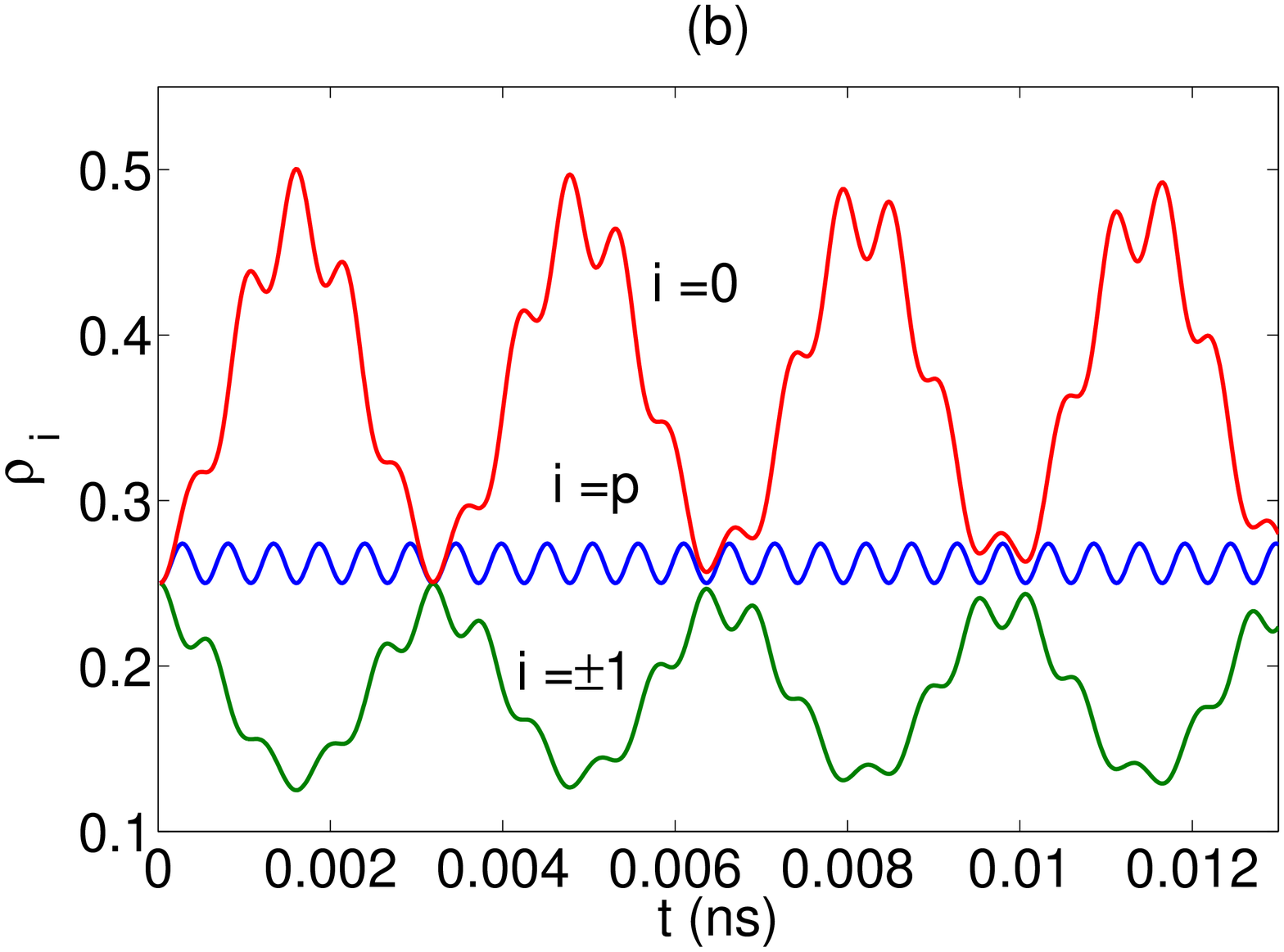}
\caption{\label{fig:GP1} Time evolution of the population fractions $\rho_i$ for a system prepared with $\rho_i(t=0)=1/4$ at densities (a) $n=10^{18}$ cm$^{-3}$ (b) $n=10^{20}$ cm$^{-3}$. In the former case, $n_{\mathrm{c}}>n$ and the spin mixing between the ortho- and para- sectors is minimal; for the latter case, $n_{\mathrm{c}}<n$ and a more substantial spin-mixing is observed. In general, spin mixing between ortho- and para- states is significant only for $n>n_{\rm{c}}$ as predicted in Eq. \ref{eq:GP2sol2}.  Note that Ps annihilation is not included in the calculations. }
\end{figure}

The internal energy difference $\epsilon$ between ortho- and para- states gives rise to a barrier for the conversion from \textit{p}-Ps to \textit{o}-Ps. This interconversion happens when the interaction that exchanges populations is sufficiently strong to overcome the energy barrier. Thus, we again encounter the competition between the $\epsilon$ and the $g_1$ terms (i.e. spin-mixing interaction) of Eq. \ref{eq:E1}. As discussed above, this competition is expressed in the comparison of the condensate density $n$ with the critical value $n_{\mathrm{c}}$, which corresponds to when the interaction energy is comparable to $\epsilon$. Using the ratio $\nu=n/n_{\mathrm{c}}=2g_1n/\epsilon$ as a parameter, we estimate the limiting behaviors of the solution Eq. \ref{eq:GP2sol2}, for which the amplitude of population variation is determined by $\rho_2-\rho_3$. When $\nu \ll 1$, $\rho_2-\rho_3\approx\nu(1-\rho(\tau_0))\rho(\tau_0)$, which is of order $\nu$. Thus, regardless of what the initial conditions are, $\rho$ is limited by this quantity, and the \textit{p}-Ps fraction will not deviate significantly from $\rho\approx \rho_3$. On the contrary, when $\nu \gg 1$, we have $\rho_2-\rho_3\approx |1-2\rho(\tau_0)|$ for $\cos2\phi(\tau_0)=1$ and $\rho_2-\rho_3\approx \rho(\tau_0)$ for $\cos 2\phi(\tau_0)=-1$, which are both of order $1$. The solution $\rho$ can take any value between 0 and 1. From this analysis, we can see the spin mixing occurs when the condensate density is greater than the critical value $n_{\mathrm{c}}$.

Taking all four states into consideration, we use Eq. \ref{eq:GP} to compute numerically the time evolution of populational fractions $\rho_i$ for $i=0,\pm1,{\rm{p}}$. The total population is conserved in this evolution, $\sum_{\substack{i}}\rho_{i}=1$. Fig. \ref{fig:GP1}(a) shows the propagation of the four components at density $10^{18}\textrm{ cm}^{-3}$, in a case for which the initial populations are all equal. The three ortho- states continuously exchange populations and exhibit sinusoidal time evolution. On the other hand, the para- component has very small oscillations, corresponding to the smallness of the para-ortho- interconversion. Since here we have $n/n_{\mathrm{c}}=\nu=0.08$, this result is consistent with the solution of the two-component system. In addition, it also implies that although the unpolarized ortho- fraction diminishes very quickly in many experimental scenarios \citep{Cassidy2010}, it would be stable in the condensed phase if $n < n_{\mathrm{c}}$. Considering the case with $n > n_{\mathrm{c}}$, we set $n=10^{20}\;\rm{cm}^{-3}$ and use the same initial conditions. The time evolutions are shown in Fig. \ref{fig:GP1}(b). Here we can clearly see a more substantial oscillatory pattern for the para- fraction and some additional small-scale fluctuations on the ortho- sector, which reveals a stronger para-ortho- spin-mixing occurring at the high density. Both cases agree very well with our former estimate regarding the critical density. 
\section{\label{sec:annihilation} Positronium annihilation and $\gamma$-ray laser\\}
Proposals for a $\gamma$-ray laser \cite{Mills2004,MillsJr1994,MillsJr2002107} call for preparing a BEC of polarized \textit{o}-Ps. This is used as a storage medium which can subsequently be changed to a gain medium by ortho-para conversion. The \textit{p}-Ps atoms can then participate in stimulated emission of the $\gamma$-rays produced in \textit{p}-Ps annihilation. As we have shown above, there are complex effects of spin mixing when the Ps density exceeds the critical density $n_{\mathrm{c}}$. We now investigate how these effects modify the population distribution within a Ps BEC that is also subject to the processes of spontaneous annihilation. We find that spin mixing can significantly modify the optimal strategy for producing a $\gamma$-ray laser.

We consider two approaches. The first is a semi-classical rate-equation approach appropriate for an incoherent mixture of Ps atoms. The second is the full solution of time-dependent GP equations, taking spontaneous annihilation into account. On the microscopic level, the GP equations exhibit behavior much different from the classical rate equations, but when spatial averaging is included, the two approaches give similar results for $n>n_{\mathrm{c}}$.

\subsection{\label{sec:rate} The rate equation approach}
In this section we use a simple rate equation approach to model Ps annihilation in arbitrary mixtures of \textit{o}-Ps and \textit{p}-Ps subject to spin-mixing collisions. This approach should be valid for incoherent Ps mixtures which can be modeled as weakly-interacting classical gases \cite{Pethick}.  It provides a reference point for understanding the dynamics of nondegenerate gases of Ps, and for comparison with the mixed Ps BECs that need to be described by the GP equations.  

Referring to the notation introduced above, the possible spin-mixing collisions can be described schematically by \citep{Ivanov2002}  
\begin{eqnarray}
\label{eq:scattering}
\left| 0 \right\rangle \left| 0 \right\rangle & \underset{k_{1}}{\overset{k'_{1}}{\leftrightharpoons }} & \left| {\rm{p}} \right\rangle \left| {\rm{p}} \right\rangle \nonumber\\
\left| +- \right\rangle & \underset{k_{2}}{\overset{k'_{2}}{\leftrightharpoons }} & \left| {\rm{p}} \right\rangle \left| {\rm{p}} \right\rangle \label{eq:Rates}\\
\left| 0 \right\rangle \left| 0 \right\rangle & \underset{k_{3}}{\overset{k'_{3}}{\leftrightharpoons }} & \left| +-\right\rangle \nonumber
\end{eqnarray}
where $k_\alpha$ and $k'_{\alpha}$ denote the rate constants for the left-to-right and inverse processes respectively, and $|+-\rangle=(|1\rangle |-1\rangle+|-1\rangle|1\rangle)/\sqrt{2}$. The first two of these processes describe direct ortho-para- interconversion. The third is associated with redistribution of ortho- state populations. Treating the inelastic scattering events as individual reactions among particles of classical gases, we can calculate the rate equations \citep{Pethick} by summing the products of rate constants, $k$,  and reactant population fractions, $\rho_i$. Adding the decay terms representing Ps annihilation, the rate equations can be expressed as 
\begin{eqnarray}
\label{eq:rate}
\dot{\rho}_{1}&=&k_{3}\rho_{0}^{2}+k'_{2}\rho_{\rm{p}}^{2}-(k_{2}+k'_{3})\rho_{1}\rho_{-1}
-\rho_{1}/{\tau_{\rm{o}}}\nonumber\\
\dot{\rho}_{0}&=&k'_{3}\rho_{1}\rho_{-1}+k'_{1}\rho_{\rm{p}}^{2}-(k_{1}+k_{3})\rho_{0}^{2}
-\rho_{0}/{\tau_{\rm{o}}}\nonumber\\
\dot{\rho}_{-1}&=&k_{3}\rho_{0}^{2}+k'_{2}\rho_{\rm{p}}^{2}-(k_{2}+k'_{3})\rho_{1}\rho_{-1}
-\rho_{-1}/{\tau_{\rm{o}}}\nonumber\\
\dot{\rho}_{{\rm{p}}}&=&k_{1}\rho_{0}^{2}+k_{2}\rho_{1}\rho_{-1}-(k'_{1}+k'_{2})\rho_{\rm{p}}^{2}
-\rho_{\rm{p}}/{\tau_{\rm{p}}}\nonumber\\
\end{eqnarray}
where $\rho_0$, $\rho_{\pm 1}$, and $\rho_{\rm{p}}$ are the fractions of the four spin states. Since the total population decays in time, we define the population fractions as $\rho_i\left(t\right)=n_i\left(t\right)/\ni$, where $\ni=n\left(t=0\right)$. We can calculate the rate constants as $k_{\alpha}=\ni \langle v \sigma_{\alpha}\rangle$, where $\langle\rangle$ stands for the thermal average over the relative velocity $v$, and the cross-sections can be calculated from first-principles quantum mechanics \cite{Ivanov2002}. The cross-sections $\sigma_1=0.130\times 10^{-14} \textrm{ cm}^2$ and $\sigma_2=\sigma_3=0.261\times 10^{-14} \textrm{ cm}^2$, corresponding to the processes in Eq. \ref{eq:scattering}, are calculated in Appendix \ref{app:Hint}. It should be noted that \textit{o}-Ps and \textit{p}-Ps have different internal energies, so that the thermal averages of the corresponding $k_{\alpha}$ and $k'_{\alpha}$ will also be different. 

\begin{figure}
 \includegraphics[width=3.3in]{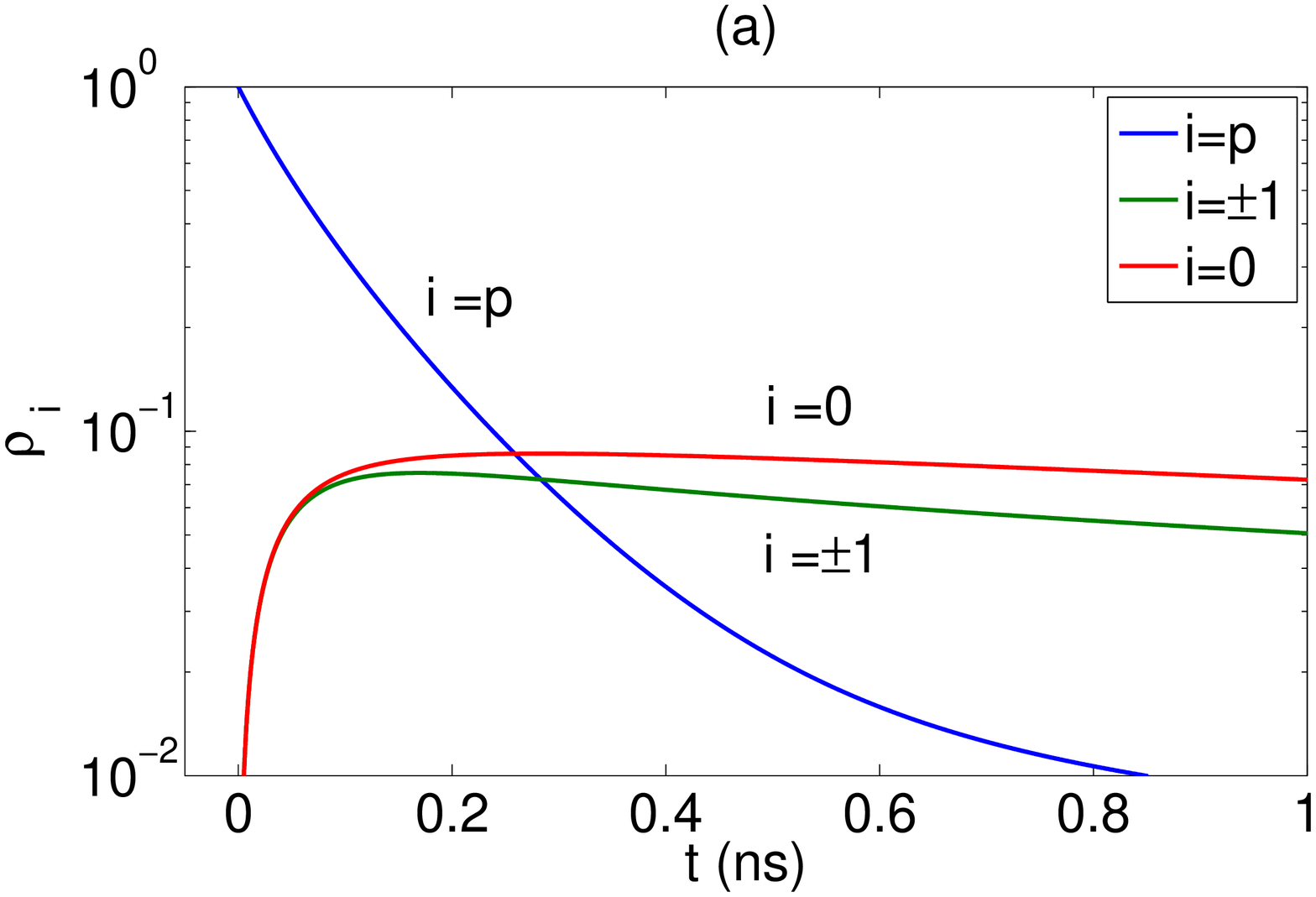}\\
 \includegraphics[width=3.3in]{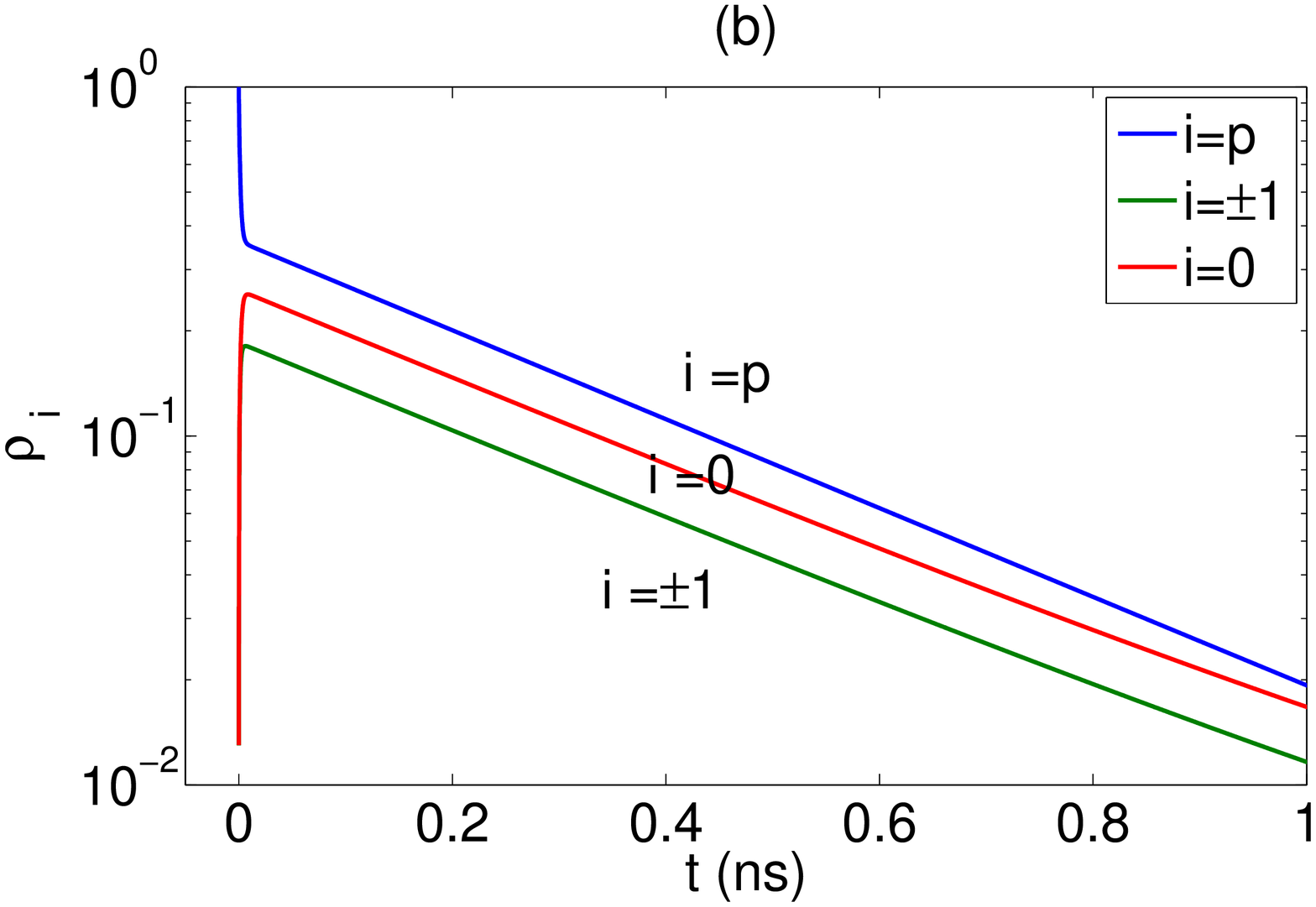}
\caption{\label{fig:rate} Evolution of the population fractions calculated with the rate equations (Eq. \ref{eq:rate}) at densities (a) $\ni=10^{18}$ cm$^{-3}$ and (b) $\ni= 10^{20}$ cm$^{-3}$. In both cases the initial population is pure \textit{p}-Ps. In the low density case, the spin mixing is not sufficiently strong to reach a quasi-equilibrium. At long times, the para- population is significantly depleted and the ortho- population slowly decays away through spin-mixing with the para- sector. In the high density case, the relative populations quickly reach a quasi-equilibrium and collectively decay with an effective time constant $\tau_{\rm{eff}}\approx 4\tau_{\rm{p}}$.}
\end{figure}

There are two main mechanisms in the rate equations. Spin mixing has the effect of distributing populations of the four states into an equilibrium, while Ps annihilation depletes populations. For the timescale of interest (1 ns), the system mainly decays through the para- state. As an example, we consider a thermal gas of Ps at temperature $T=15\,K$ and density $\ni=10^{18}$ cm$^{-3}$. For these conditions, the magnitudes of the various rate constants are between $10^{9}\textrm{ s}^{-1}$ and $10^{10}\textrm{ s}^{-1}$, which is comparable to the \textit{p}-Ps decay rate. We propagate the rate equations in Eq. \ref{eq:rate} with initial condition $\rho_{\rm{p}}(t=0)=1,\;\rho_{\rm{o}}(t=0)=0$ and plot the resulting time evolution in Fig. \ref{fig:rate}(a). Initially a small portion of \textit{p}-Ps is converted to \textit{o}-Ps. After \textit{p}-Ps decays below a threshold $k_{1,2}\rho_{\rm{p}}\lesssim 1/\tau_{{\rm{p}}}$ further conversion from para- to ortho- becomes small and the ortho- population begins to decay away. When the para- state is mostly depleted, the dominant decay process comes from conversion from ortho- to para- followed by para- self-annihilation. At this point, the ortho-to-para spin-mixing rate is slower than the \textit{p}-Ps decay rate, which creates a bottleneck that slows down the decay process. Thus, at $t=1$ ns, we can still see a fraction of the total population remains in the ortho- sector. 

At a density of $ \ni=10^{20}\;{\rm{cm}}^{-3}$, all the rate constants are increased by two orders of magnitude. The same initial conditions give rise to a qualitatively different evolution as shown in Fig. \ref{fig:rate}(b). Since the spin mixing rates are much stronger than the \textit{p}-Ps decay rate, we expect rapid population redistribution on a timescale shorter than $\tau_{\rm{p}}$. Thus the relative populations remain in quasi-equilibrium at all times during the decay process. As shown in Fig. \ref{fig:rate}(b), all components decay with an effective lifetime $\tau_{\rm{eff}}\approx4\;\tau_{\rm{p}}$ during this quasi-equilibrium redistribution. For the same reason, regardless of the initial state of the system, it quickly acquires some mixture of the para- state and decays, so that the system loses its population more rapidly than for the lower density case. By $t=1$ ns, a substantial fraction of initial Ps has decayed; at longer times, the population decays more slowly due to the bottleneck effect  discussed above. 

\subsection{\label{sec:GPsimulation} The Gross-Pitaevskii equations}

In this section, we follow the GP formulation provided in Sec. \ref{sec:GP} and take into account the effect of Ps annihilation. To incorporate the Ps decay into the formulation, we use an effective Hamiltonian which modifies the internal energies of individual species with an imaginary component
\begin{eqnarray}
\label{eq:decay}
\epsilon_{\rm{o}}&=&\epsilon-i\frac{\hbar}{2\tau_{\rm{o}}},\nonumber\\
\epsilon_{\rm{p}}&=&-i\frac{\hbar}{2\tau_{\rm{p}}}.
\end{eqnarray}
Using the internal energies in Eq. \ref{eq:decay} and neglecting the spatial dependence in Eq. \ref{eq:GP}, we calculate the time evolution of a condensate prepared in the para- state at various densities. As discussed in Sec. \ref{sec:GP}, when $\ni>n_{\mathrm{c}}$, the initial \textit{p}-Ps condensate is no longer a ground state and the system will undergo strong spin-mixing. 

We demonstrate this by considering a system prepared in a predominantly para- state, with a small admixture of ortho-.  In particular, our initial state consists of equal populations of the three ortho- states, each with population fraction $\sim 10^{-7}$. The phases were determined using imaginary time evolution to find the lowest energy state consistent with this population distribution. We then numerically integrate Eqs. \ref{eq:GP} starting with this initial state.

\begin{figure} [t]
 \includegraphics[width=3.3in]{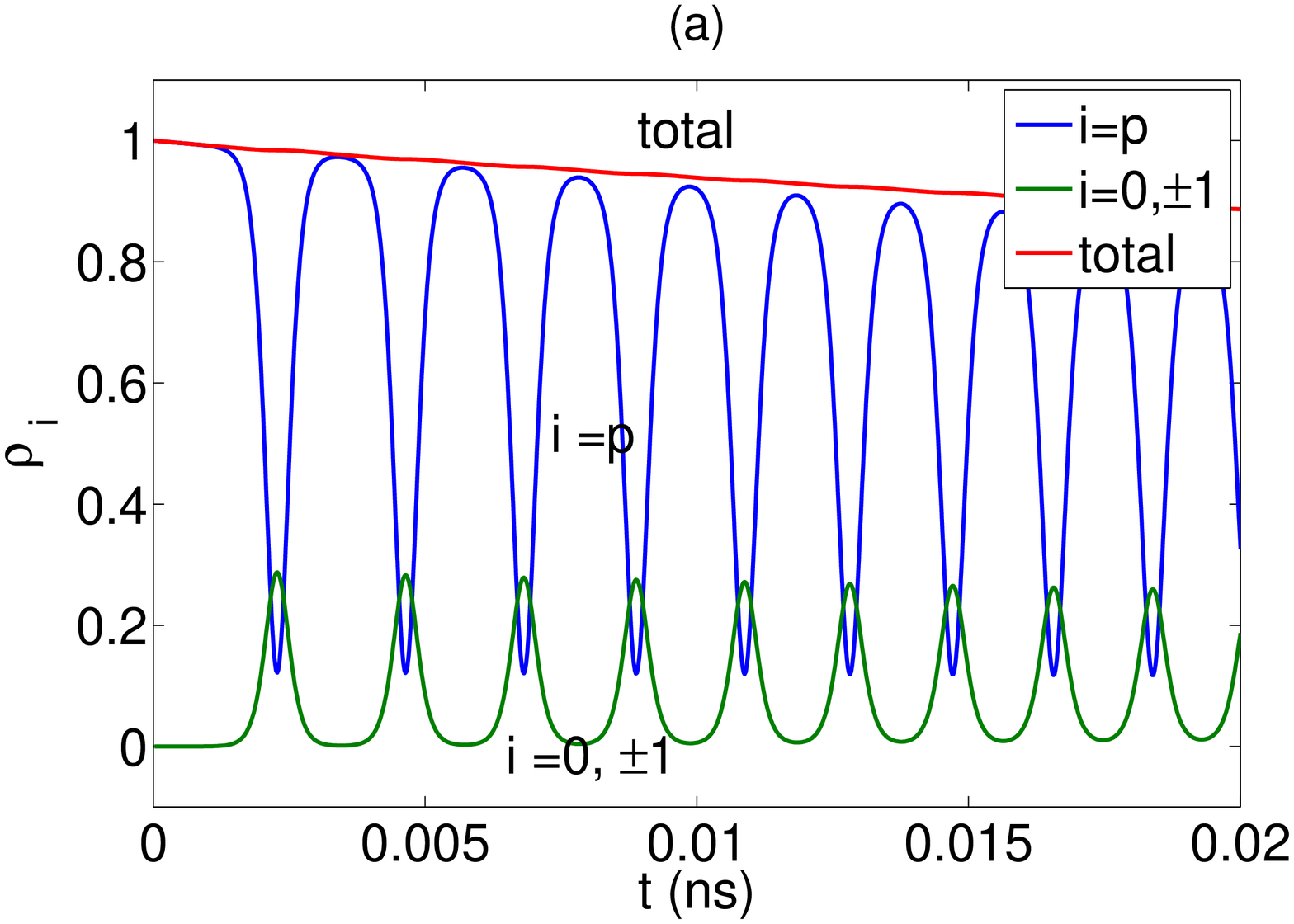}\\
 \includegraphics[width=3.3in]{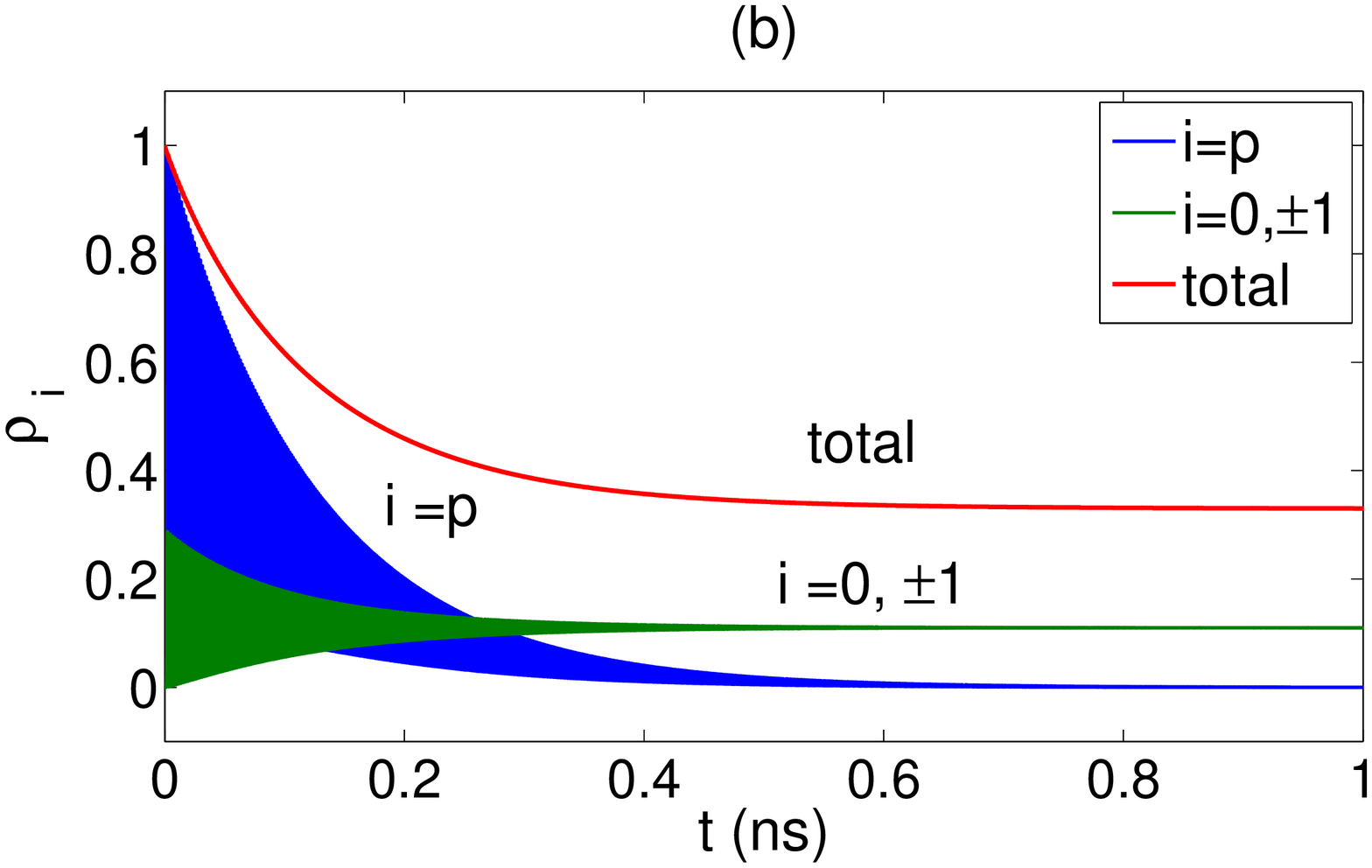}
\caption{\label{fig:GP2} Population evolution of a mixed Ps condensate undergoing both spin mixing and Ps annihilation over timescales of (a) $0.02 \;\rm{ns}$ and (b) $1 \;\rm{ns}$. The system is prepared in $|{\rm{p}}\rangle$ at a density of $\ni=10^{20}$ cm$^{-3}$. The short-scale evolution shows rapid oscillations due to spin mixing. The long-time evolution reveals a prominent decay behavior, but with about one third of the total population trapped in the ortho- sector. }
\end{figure}

Figs. \ref{fig:GP2} (a) and (b) show the time evolution of a condensate at $\ni=10^{20}$ cm$^{-3}$ at different time scales. Figure (a), which depicts a shorter timescale, reveals a quasi-periodic oscillation pattern caused by the spin-mixing. As time progresses, the amplitude of the oscillations decreases due to the decay term introduced in the GP equation. As $t$ becomes comparable to $\tau_{\rm{p}}$, both the para- population and its population exchange with the ortho- state begin to disappear. However, the total population does not simply decay away through the para- state. It evolves into a pure \textit{o}-Ps state that is decoupled from the para- annihilation process. About one third of the population is trapped in the ortho- sector as $t>1 \,\rm{ns}$. This behavior is qualitatively different from that observed in the rate-equation approach, where there is no phase coherence of population amplitudes. In contrast to the classical rate equations, in the GP equation, the spin-mixing interactions depend on both the reactant and product densities, and ortho--para conversion is suppressed if either density is small \cite{Heinzen2000}.

In the calculations discussed above, we assumed all components of the condensate occupy the same homogeneous spatial profile. We now discuss more realistic cases, in which this assumption is not made. First we consider a one-dimensional quasi-homogeneous condensate, in which each wavefunction is modulated at $t=0$ with random spatial noises. 

In particular, we assume the transverse directions are tightly confined in a box smaller than the healing length, $\xi$. We then define a new 1D pseudo-potential constant such that the interaction energy per particle is the same as a 3D system with density $\ni$ after integrating over the transverse directions. We prepare an initial state that corresponds to the state chosen above, defined on a uniform grid of 1000 points. Then, at each grid point $m$ we multiply the wave function of each component $i$ by a factor of $1 + \eta_{m,i}$, where $\eta_{m,i}$ is a complex Gaussian random number with $\langle \eta_{m,i}^* \eta_{m^\prime,i^\prime} \rangle = 2 \sigma^2 \delta_{m,m^\prime} \delta_{i,i^\prime}$, $\langle \eta_{m,i} \eta_{m^\prime,i^\prime} \rangle = \langle \eta_{m,i} \rangle = 0$, and $\sigma = 0.01$.
We emphasize that this propagation is not intended to simulate the effects of dissipation beyond that associated with electron-positron annihilation. It amounts to applying a small relative change to the populations of the initial state discussed above, in order to test the sensitivity of time evolution to initial conditions.

\begin{figure} [t]
 \includegraphics[width=3.3in]{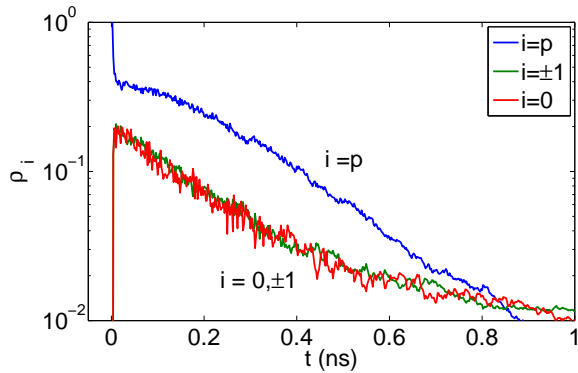}
\caption{\label{fig:noise} Time evolution of a mixed condensate of initial density $\ni=10^{20}$ cm$^{-3}$ prepared in the para- state with random spatial noise. At short times, spin-mixing drives the populations toward the equilibrium distribution. Then, all species decay exponentially at the same rate, given approximately by $4 \tau_p$, subject to background fluctuations. Due to the presence of noise, no population is trapped in the ortho- sector. The population evolution is qualitatively similar to that obtained from the rate equations (see Fig. \ref{fig:rate} (b)) with fitted decay lifetimes equal to $\tau_{\rm{eff}}/\tau_{\mathrm{p}}=$ 4.3, 4.3, 4.3, 3.7 for states $i=1,0,-1, \mathrm{p}$, respectively.}
\end{figure}

\begin{figure} [t]
 \includegraphics[width=3.3in]{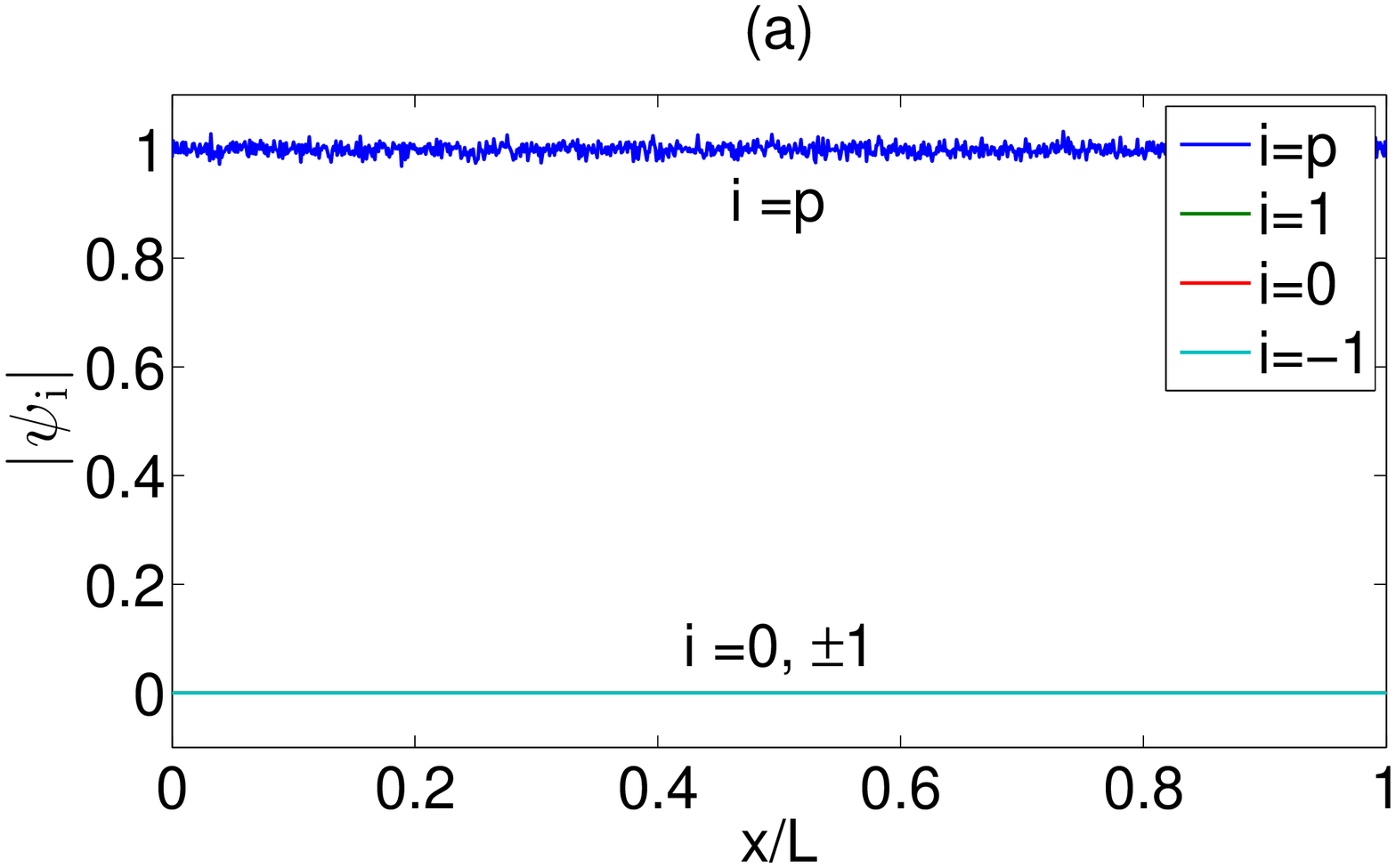}\\
 \includegraphics[width=3.3in]{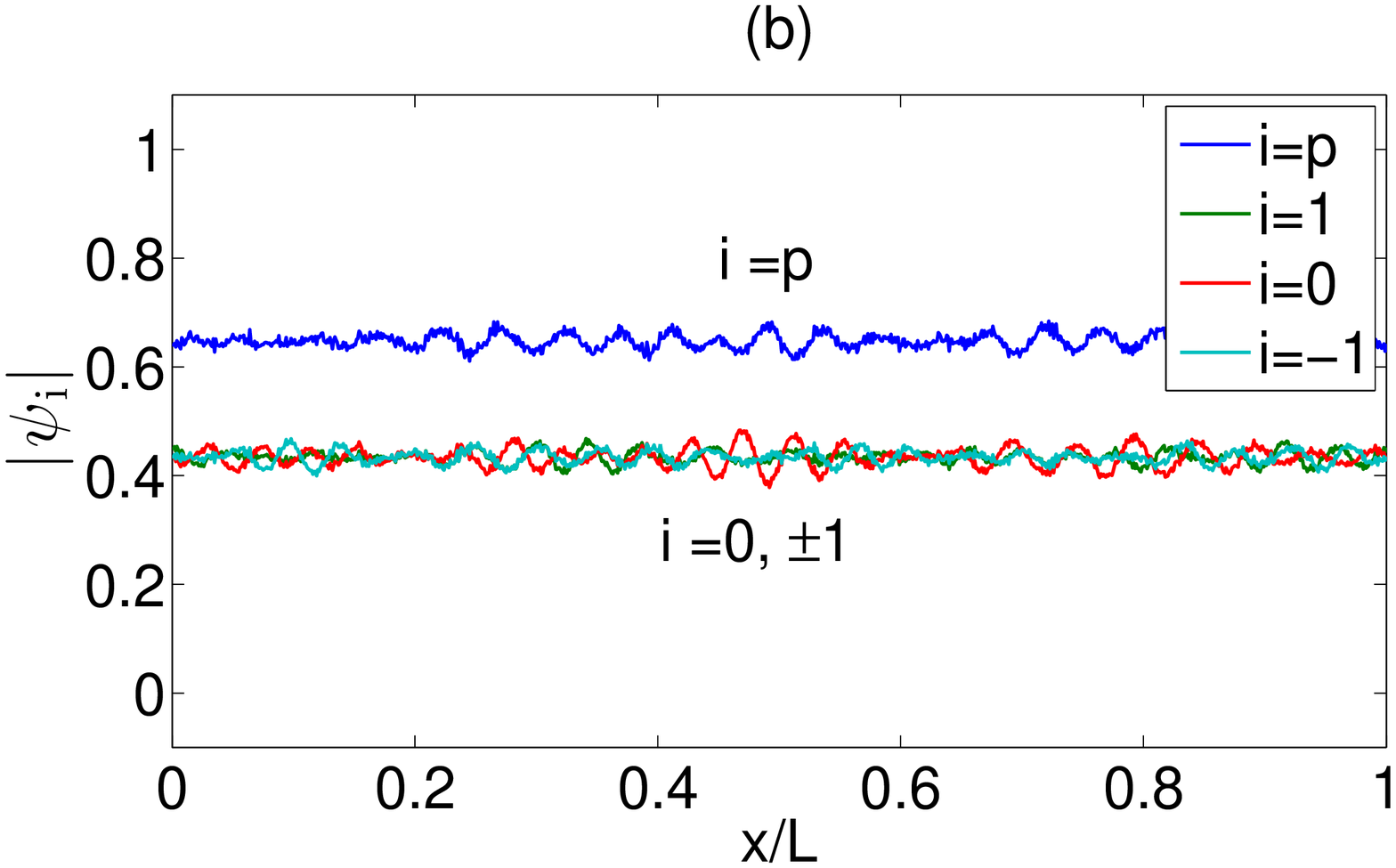}\\
 \includegraphics[width=3.3in]{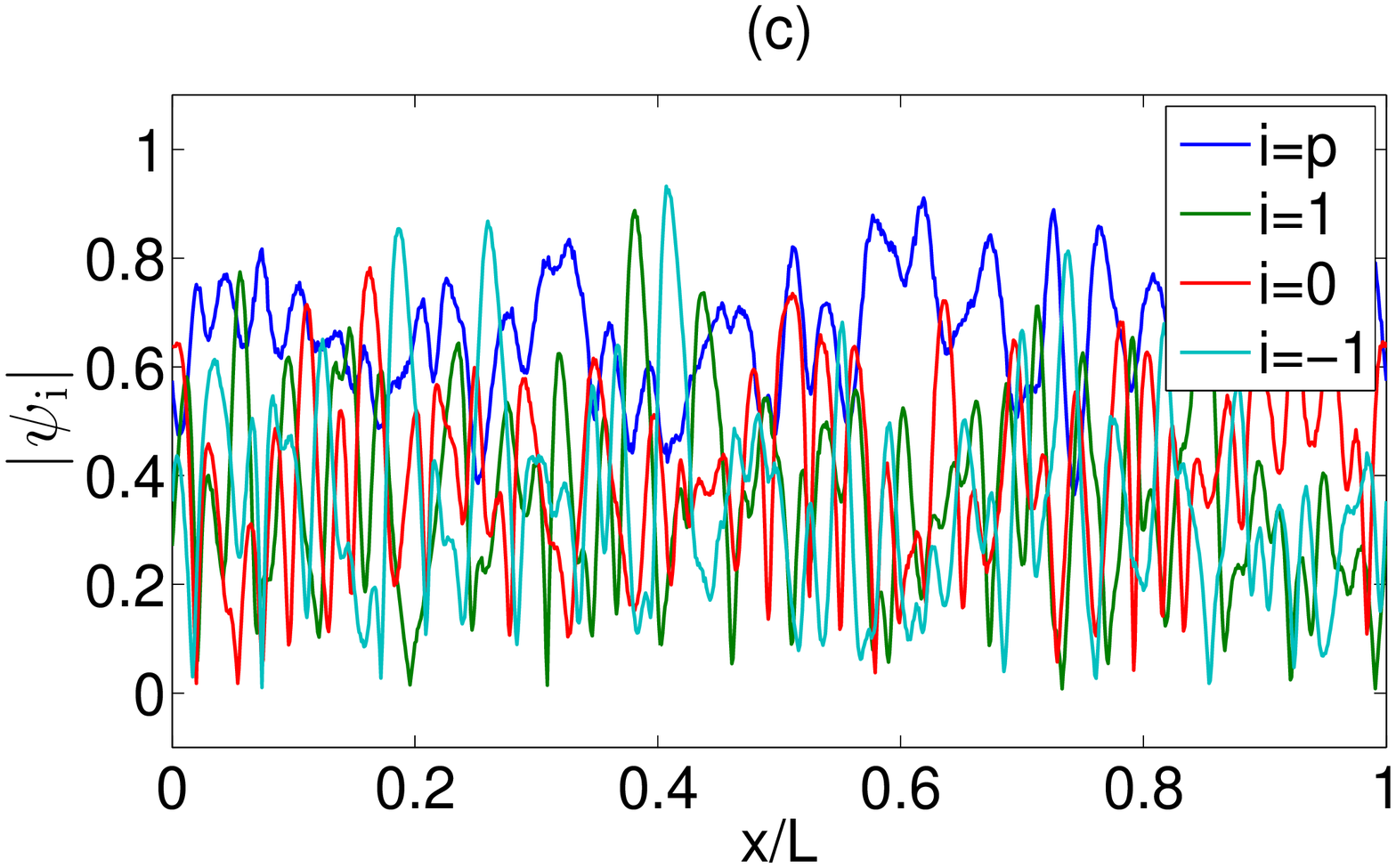}
\caption{\label{fig:noise2} The wavefunction amplitudes $\left|\psi_i(x)\right|$ recorded at (a) $t=0$ ns, (b) $t=0.002$ ns and (c) $t=0.004$ ns for the evolution in Fig. \ref{fig:noise}. The system size $L$ is selected such that $L\sim 200\xi$. The noise grows significantly within one cycle of spin-mixing oscillation. At $t=0$ ns, we introduce small random noises to each of the initial wavefunctions $\psi_i(x)$. At $t=0.002$ ns, $\left|\psi_i(x)\right|$ follows the evolution pattern of Fig. \ref{fig:GP2} and is about uniform with small spatial fluctuations. At $t=0.004$ ns, the background fluctuations begin to dominate and destroy the uniform spatial structures. By this time, the irregular spin structures have formed such that the system undergoes local spin mixing.}
\end{figure}

We find that when $\ni>n_{\mathrm{c}}$, the initial spatial noise grows rapidly in time into a regime of highly irregular evolution.
For example, we consider the evolution of a condensate as prepared above, at a density of $\ni=10^{20}\mathrm{ cm}^{-3}$, as shown in Fig. \ref{fig:noise}. All species decay with about the same time constant $\tau \approx 4 \tau_p$, and there is no persistence of a trapped \textit{o}-Ps population that occurs in the absence of noise. 
In this regime, spin structures of size comparable to the healing length rapidly collide and inter-convert. In Fig. \ref{fig:noise2}, we can see the noise grows substantially during one cycle of spin mixing oscillation, giving rise to irregular spatial structure.
This irregular spatial evolution guarantees that there are always regions where \textit{o}-Ps can be converted to \textit{p}-Ps, and then decay. 
As suggested by Fig. \ref{fig:noise}, the evolution of the relative populations is qualitatively similar to that found in the rate equation approach. The overall density also decays on a timescale set by $\sim 4 \tau_{\rm p}$ instead of $\tau_{\rm p}$. This is because the irregular spatial variation plays a role similar to that of phase averaging in an incoherent mixture.

\begin{figure} [t]
 \includegraphics[width=\columnwidth]{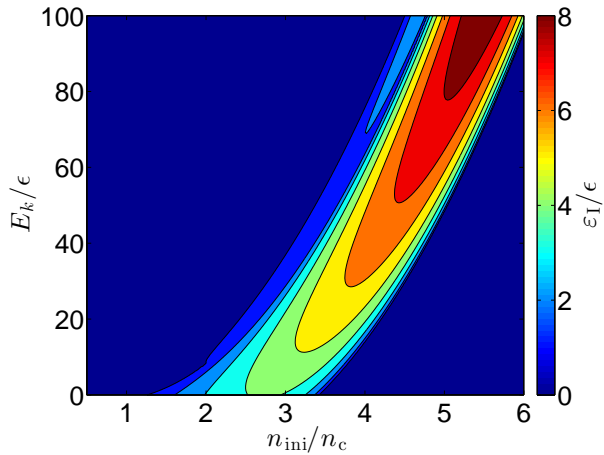}
\caption{\label{fig:BdG} Imaginary part of the energy, $\varepsilon_{\rm I}$, of the most unstable BdG mode as a function of condensate density $\ni$ and kinetic energy $E_k$. We express these variables in dimensionless form as indicated by the axis labels. For $\ni<n_{\mathrm{c}}$, all BdG modes are stable, i.e. the imaginary part vanishes. When $\ni$ increases above $n_{\mathrm{c}}$, the low energy (long wavelength) modes become unstable. For any value of $\ni>n_{\mathrm{c}}$, there is an energy band of instability, whose width is about four orders of magnitudes greater than the natural $\gamma$-ray linewidth ($\Delta E\sim 10^4\hbar/\tau_{\rm{p}}$).}
\end{figure}

The short-time amplification of the noise in the GP evolution can be understood by solving the Bogoliubov-de Gennes (BdG) equations \cite{Kawaguchi2012253}, as shown in Appendix \ref{app:BdG}. We find that when $n>n_{\mathrm{c}}$, a condensate initially in the para- state has complex BdG eigenvalues, $\varepsilon=\varepsilon_{\rm{R}}+i\varepsilon_{\rm{I}}$.
If $\varepsilon_{\rm{I}}\neq 0$, the mode amplitude will grow exponentially until the nonlinear terms in the GP equations dominate. This regime is characterized by significant spatial irregularity on a short length scale. The condensate therefore is composed of many spatial modes, and correspondingly has a wide distribution in momentum space. This may reduce the $\gamma$-ray yield, as studied in \cite{Mills2004}. We characterize the instability of a condensate by the largest value of $\varepsilon_{\rm{I}}$ of any of its BdG modes, $\varepsilon_{\rm{I}}^{\rm{max}}$. Fig. \ref{fig:BdG} shows the dependence of $\varepsilon_{\rm{I}}^{\rm{max}}$ as a function of kinetic energy and \textit{p}-Ps density. It shows that the unstable modes appear when $\ni>n_{\mathrm{c}}$, which is the same condition for the occurrence of spin mixing. Thus, when spin mixing is most pronounced, there is also a propensity for spatial inhomogeneity of the system.

\subsection{Optimizing $\gamma$-ray yield}

\newcommand{\omegarf}{\omega_0}

It has been shown that \cite{Liang1988} the formation of a Ps BEC corresponds to large \textit{p}-Ps stimulated annihilation cross-section. For zero temperature BECs, a macroscopic number of Ps atoms occupy the zero-momentum mode. This results in a small energy uncertainty and a narrow $\gamma$-ray line-width determined by the natural lifetime: $\Delta E  \sim \hbar/\tau_{\mathrm{p}} = (1/2)\;\alpha^5m_{\mathrm{e}}c^2$, where $\alpha$ is the fine structure constant \cite{Mills2004}. As the population of states in $\mathbf{p}\neq 0$ modes increases, both the energy uncertainty and the $\gamma$-ray line-width will increase. This will decrease the resulting coherent $\gamma$-ray yield. As discussed in the previous section, the instability associated with the spin mixing would lead to a broad uncertainty in kinetic energies that is about four orders of magnitude greater than the natural $\gamma$-ray linewidth (see Fig. \ref{fig:BdG}). The corresponding $\gamma$-ray yield would be greatly reduced. To avoid this effect, it is crucial to conduct the experiment such that the condensate is composed primarily of the zero momentum mode.

Original proposals suggested that lasing can be initiated by converting an ortho- condensate into a para- condensate by application of a radio frequency pulse \cite{MillsJr2002107}. For dense condensates, this procedure could result in a spread of momentum states due to the spatial instability demonstrated above. However, if the para- fraction is kept sufficiently small by properly manipulating the duration and strength of external pulses, this instability can be avoided, and the $\gamma$-ray yield maximized. In the following, we model a polarized Ps BEC subject to external magnetic fields. We provide two ideal schemes for ortho- to para- conversion, and subsequent $\gamma$-ray emission, without exciting spin-mixing instabilities. To calculate the final gain of a $\gamma$-ray laser, it is necessary to consider the actual construction of a laser, the geometry structure of the cavity, and how the stimulated annihilation is initiated with respect to that structure. Future simulations can be made with the inclusion of these factors.

We now consider the effect of an external magnetic field ${\bf B} = (B_x, B_y, B_z)$ applied to a Ps atom. Using the same basis as Eq. \ref{eq:GP}, the atom-field interaction is given by
\begin{eqnarray}
\label{eq:H_ext}
H_{\rm{ext}}= -\mu_{\rm e}
\left( \begin{array}{cccc}
  0 & 0 & 0 & \frac{-B_x+iB_y}{\sqrt{2}} \\ 
  0 & 0 & 0 & B_z \\ 
  0 & 0 & 0 & \frac{B_x+iB_y}{\sqrt{2}} \\ 
  \frac{-B_x-iB_y}{\sqrt{2}} & B_z & \frac{B_x-iB_y}{\sqrt{2}} & 0 
\end{array}\right),\nonumber\\  
\end{eqnarray} 
where $\mu_{\rm e}$ is the magnetic moment of the electron \cite{RMP2012}.
This matrix shows that a field in either the $\hat{x}$ or $\hat{y}$ direction couples the $|\pm 1\rangle$ states with  $|\mathrm{p}\rangle$ with equal probability. A field in the $\hat{z}$ direction induces coupling only between $|0\rangle$ and  $|{\rm{p}}\rangle$. In order to couple the state $|\pm 1\rangle$ to the state $|\mathrm{p}\rangle$, we can use circularly polarized light. We assume an rf pulse where $B_x = B \cos(\omegarf t)$, $B_y = \pm B \sin(\omegarf t)$, and $B_z=0$, where $\omegarf$ is the frequency of the rf field and $+(-)$ denotes right(left)-hand polarization. If $\hbar \omegarf \sim \epsilon$, we can invoke the rotating wave approximation, and eliminate the counter rotating terms. In this approximation, left(right)-handed polarization will couple only the state $|1\rangle$ ($|-1\rangle$) to the state $|{\rm{p}}\rangle$. Note that if linear polarization were used, i.e., $B_x = B \cos(\omegarf t)$ and $B_y=B_z=0$, then half of the population would be trapped in a ``dark state'' of the coupling Hamiltonian, and at most half of the ortho- states would be converted to para-.

Below the critical density, a strong $\gamma$-ray beam can be created by converting the entire population from $|1\rangle$ to $|\mathrm{p}\rangle$ with a single strong circularly polarized pulse. However, for $\ni > n_{\mathrm{c}}$, care must be taken to avoid the spin-mixing instabilities discussed in the previous section. An instability can occur if a  sufficiently large population is converted to the para- state. Therefore, it is necessary to design a pulse sequence that keeps the para- fraction small. 

To do this, note that for sufficiently small para- population, spin-mixing effects are frozen out and the spin evolution will be dominated by the Rabi oscillations induced by the rf pulse. To choose an appropriate pulse sequence, we consider a homogeneous two state model, consisting of only $|1\rangle$ and $|\mathrm{p}\rangle$, and the density-density terms. The effective GP equation in the rotating frame is then
\begin{eqnarray}
\label{eq:field}
i \partial_t \begin{pmatrix} \psi_1 \\ \psi_\mathrm{p} \end{pmatrix} = 
\begin{pmatrix}
0 & \Omega \\
\Omega & -i \gamma + \delta +  g_1 n_\mathrm{p} / \hbar
\end{pmatrix}
\begin{pmatrix} \psi_1 \\ \psi_\mathrm{p} \end{pmatrix} \label{eq:EigVal}
\end{eqnarray}
where $\Omega = -\mu_e B / \hbar$, $\gamma = 1 / {2 \tau_{\rm{p}}}$, $\hbar \delta = \epsilon - \hbar \omegarf$ and we have neglected the finite ortho- lifetime since it will not be relevant on the timescales of the system. In what follows, we assume that we are on rf resonance, i.e., $\delta = 0$, unless otherwise stated. 

\begin{figure} [t]
 \includegraphics[width=3.3in]{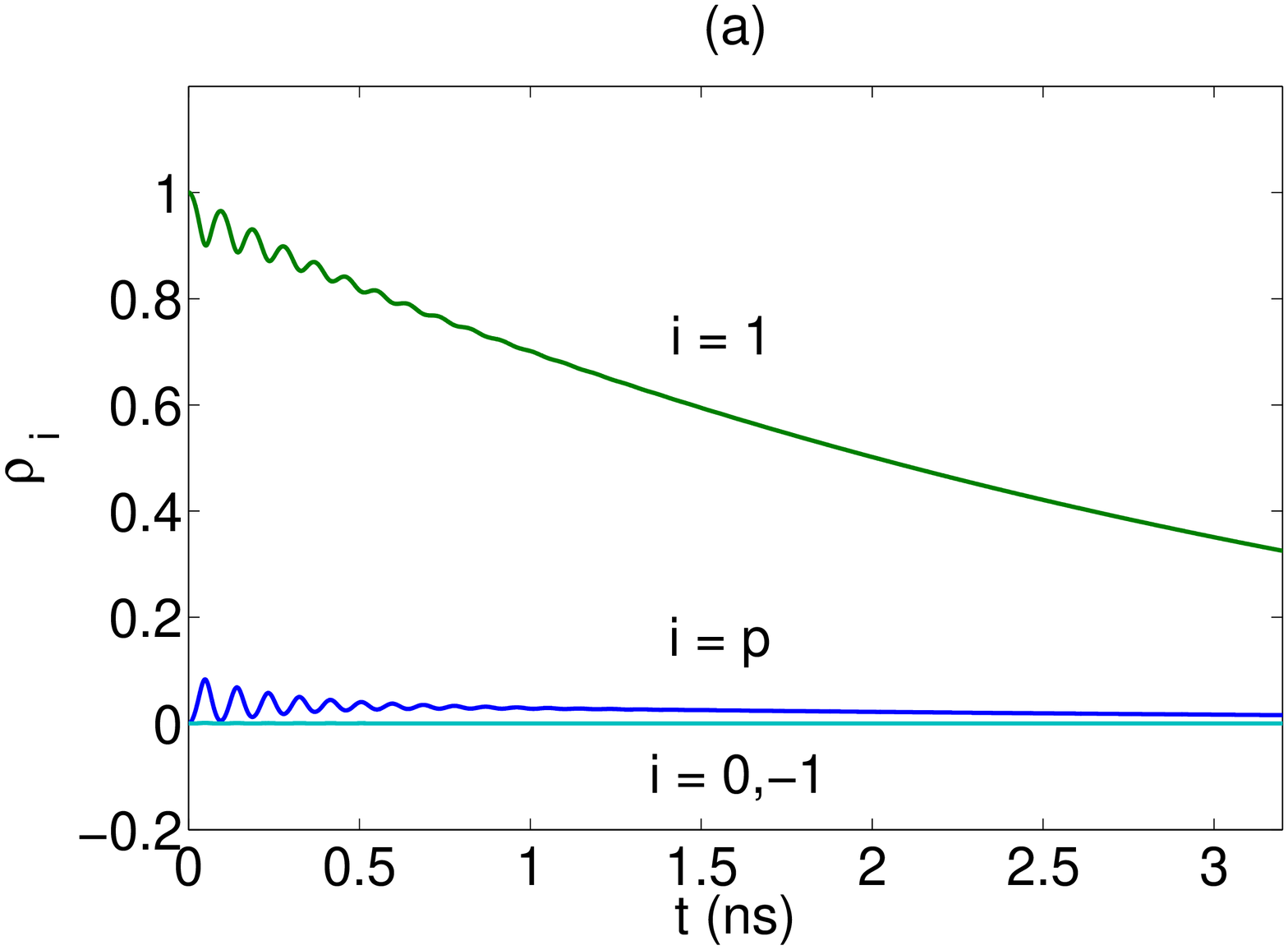}\\
 \includegraphics[width=3.3in]{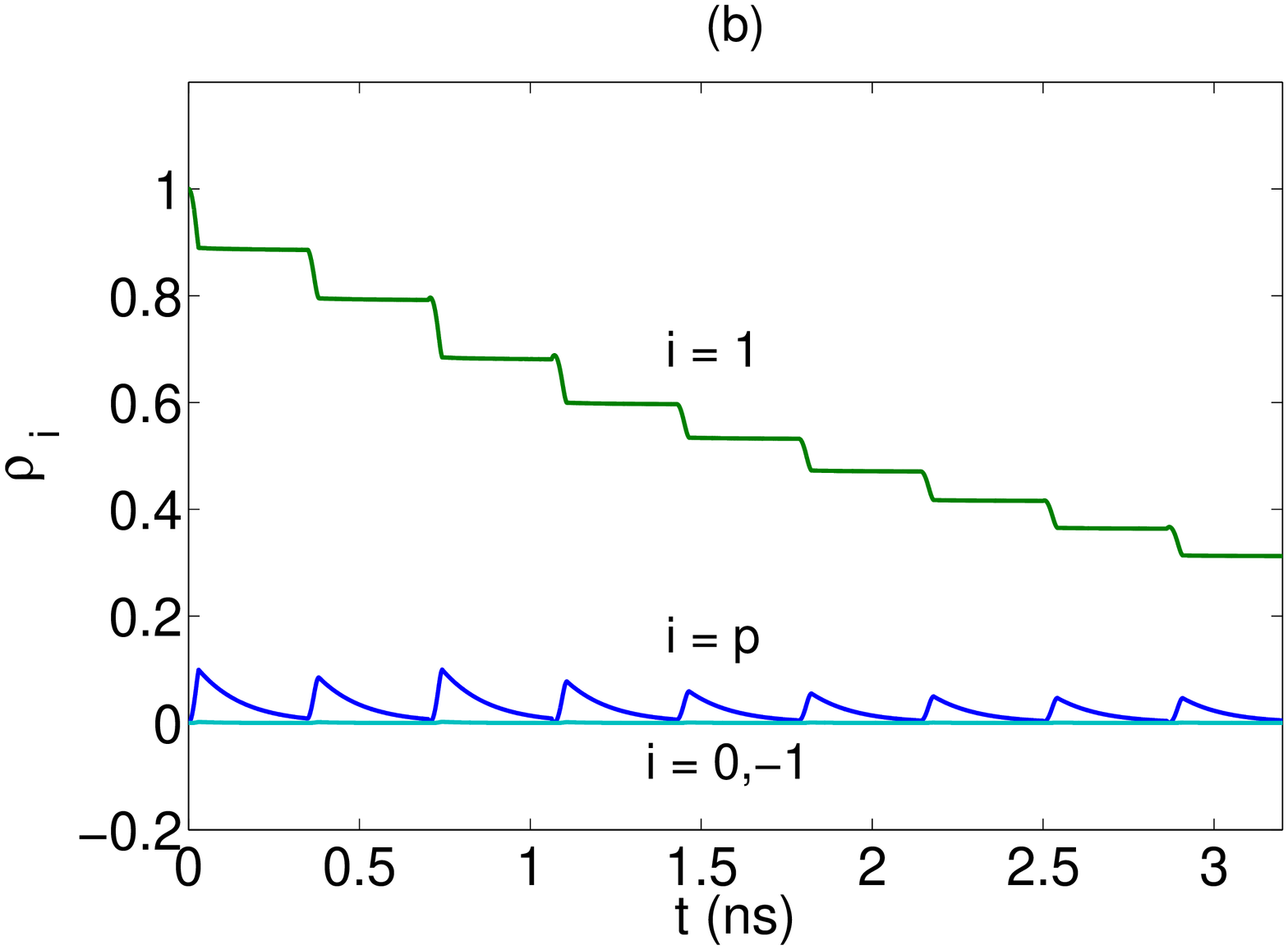}
\caption{\label{fig:GP3} Time evolution of a polarized Ps condensate of density $10^{20}\mathrm{ cm}^{-3}$ (a) under a single circularly-polarized rf pulse of frequency $\omega= \epsilon/\hbar$ and field strength $B=1/\hbar\mu_{\rm e}\tau_p$ (b) under a series of circularly-polarized rf pulses of the same frequency but with field strength $B=0.1\;{\rm{T}}$. In both cases, Ps atoms are transferred from $|1\rangle$ to $|{\rm{p}}\rangle$ using $B$ fields that are chosen to restrict the maximal $n_{\rm{p}}$ to be less than $n_{\rm c} \approx 0.12 \ni$. Since $n_{\rm{p}} < n_{\rm{c}}$, spin-mixing and the corresponding instability do not appear throughout the process.}
\end{figure}

We first ignore the effects of interactions by setting $g_1=0$, and express Eq. \ref{eq:EigVal} as an eigenvalue problem by assuming time dependence of the form $\psi \sim e^{-i \omega_\pm t} \chi_\pm$. The solution is characterized by two eigenmodes $\chi_\pm$ with eigenvalues $ \omega_\pm = \pm \omega - i\gamma/2 $, where $\omega = \sqrt{ \Omega^2 - \left(\gamma/2\right)^2 }$. The real part of $ \omega_\pm$ is responsible for the oscillatory behavior of the modes, and the imaginary part contributes to the exponential decay. Solving for the initial conditions $\psi_1(0)= \sqrt{\ni}$ and $\psi_{\rm{p}}(0) = 0$, the para- population is given by 
\begin{equation}
n_{\rm{p}} (t) = \ni e^{-\gamma t} \sin^2 \left( \omega t \right) \Omega^2 / \omega^2. \label{eq:Underdamp}
\end{equation} 
The evolution is characterized by two regimes. In the under-damped regime, where $\Omega > \gamma/2$ and $\omega$ is real, population can oscillate between $|1\rangle$ to $|\mathrm{p}\rangle$ with frequency $\omega$ and an overall decay rate $\gamma$. In the overdamped regime where $\Omega < \gamma/2$, $\omega$ becomes imaginary and Eq. \ref{eq:Underdamp} can be equivalently expressed as $n_{\rm{p}} (t) = \ni e^{-\gamma t} \sinh^2 \left( |\omega| t \right) \Omega^2 / |\omega|^2$.
The para- fraction rises to a maximum value of $n_{\rm{p}} = \ni \exp[{- \gamma / |\omega| \cosh^{-1}( \gamma / 2 \Omega ) }]$ before decaying away exponentially with slower decay rate given approximately by $ \gamma-2\mathrm{Im}[\omega]$. 

If nonlinear interactions are included, i.e., $g_1 \neq 0$, we find small oscillations in either regime. These can be understood by replacing the interaction term $g_1 n_{\rm{p}}$ with an effective detuning $\delta_{\rm eff} = g_1 \bar{n}_{\rm{p}} / \hbar$, where $\bar{n}_{\rm{p}}$ is the para- density time averaged over the short oscillations. With this replacement, the eigenvalues gain a real component which allows the corresponding modes to oscillate. At longer times, this mode damps out and the long-time behavior is well approximated by the non-interacting model. Note that these oscillations can be reduced by choosing the detuning such that $\delta \approx - \delta_{\rm eff}$.

To test the efficacy of this pulse sequence, we perform a simulation using the previously described quasi-1D method applied to a condensate initially prepared in $|1\rangle$ at a density of $\ni=10^{20}\textrm{ cm}^{-3}$. The circularly polarized rf field is added according to Eq. \ref{eq:H_ext}, without invoking the rotating wave approximation. A small random noise is added as in the previous section. Two pulse forms are considered. In Fig. \ref{fig:GP3}(a), a continuous rf pulse is applied such that the maximum para- population is set at about $n_{\mathrm{c}}$. In this case the solution of Eq. \ref{eq:field} is in the under-damped regime, in which the para- fraction grows, and then decays as described above. Rapid oscillations are present at short times before quickly decaying away. In Fig. \ref{fig:GP3}(b), a series of short pulses are applied to pump the para- population. Between every two pulses is a period of no rf fields during which the system undergoes coherent decay. The field strength is 0.1 T as suggested in \cite{MillsJr2002107} and the pulse durations are selected such that $n_{\rm{p}}$ does not exceed $n_{\mathrm{c}}$. In both pulse sequences, the absolute density of the para- condensate is kept below the critical density, and no spin-mixing instabilities form. This comes at the cost of a longer overall decay time. 

\section{\label{sec:conclusion} Conclusions}

In summary, we have considered the effects of interactions in a BEC of positronium atoms. We first derived the many-body interaction Hamiltonian describing the interactions of the ortho- (triplet) and para- (singlet) sectors. We found that the interaction Hamiltonian has an $O(4)$ symmetry. The ortho-para- energy difference breaks this symmetry to $SO(3)$, that of the triplet sector. We calculate the ground state for a uniform, homogeneous system. Below the critical density $n_{\mathrm{c}} = \epsilon / 2 g_1 \approx 1.2 \times 10^{19} \textrm{ cm}^{-3}$, the ground state is entirely polarized in the para- state. Above the critical density, the ground state contains a non-zero fraction of the ortho- state. We then consider the effects of interactions on spin dynamics for a uniform system prepared in a non-stationary state. We find that the critical density again characterizes the density for which spin mixing between the ortho- and para- sectors becomes significant.

We then consider how the effects of spin mixing will affect the use of a Ps BEC as a $\gamma$-ray laser. We develop a rate equation approach to describe an incoherent thermal mixture, including the effects of spontaneous electron-positron annihilation. When the spin mixing is strong the entire system decays with a lifetime approximately four times the lifetime of the para- state. We then model a coherent condensate using a 1D GP equation, and find that there are significant spin-mixing effects above the critical density. This spin mixing induces high frequency spatial modulations, which have the effect of averaging over populations in a way that resembles phase averaging in an incoherent system. The evolution and decay of the populations is qualitatively similar to the rate equation approach. Furthermore, the large spread of kinetic energies of the cloud will reduce the yield of a $\gamma$-ray laser. This suggests that a straightforward preparation of a high density para- condensate will not result in an optimal $\gamma$-ray yield.

Finally, we consider an experimentally relevant setup where an initially spin-polarized beam of positrons is used to prepare a Ps BEC that is initially spin polarized in the $|1\rangle$ state. We consider the effect of an external rf field used to convert the initial state to the para- state for the production of coherent $\gamma$ rays. For a circularly polarized field with frequency that is nearly resonant with the internal energy splitting of the ortho- and para- states, the ortho- condensate can be fully converted to the para- condensate. If the initial density is above $n_{\mathrm{c}}$, a modified pulse sequence can be used to avoid significant para- population, at the cost of a longer overall timescale. This suggests that for large densities, the $\gamma$-ray yield is optimized by transferring population more slowly. This comes at the cost of a lower $\gamma$-ray peak amplitude, but spread over a longer time. 

\section{Acknowledgements}

We thank David B. Cassidy for bringing this problem to our attention and for helpful comments and suggestions. We acknowledge the financial support by the NSF through the Physics Frontier Center at JQI, and the ARO with funds from both the Atomtronics MURI and DARPA's OLE Program.

\appendix
\section{Derivation of $\mathcal{H}_{\rm{int}}$ and scattering cross-section}
\label{app:Hint}
We now derive the many-body interaction Hamiltonian for a dilute Ps gas. For a sufficiently cold and dilute gas, two Ps atoms can only interact through an overall $s$-wave interaction. Expressing their interaction in a basis of $e^{-}$ and $e^{+}$ pairs, we have
\begin{eqnarray}
\label{eq:U}
U(\mathbf{r},\mathbf{r}^\prime)&&=\delta(\mathbf{r}-\mathbf{r}^\prime)\nonumber\\
&&\times\sum^{\substack{s^{+}=s^{-}}}_{\substack{s^{+}m^{+}\\s^{-}m^{-}}}U_{s}
|s^{+}m^{+};s^{-}m^{-}\rangle \langle s^{+}m^{+};s^{-}m^{-}|,\nonumber\\
\end{eqnarray}
where
\begin{eqnarray}
U_{s}=\frac{4\pi\hbar^{2}a_{s}}{m}
\end{eqnarray}
is the effective interaction expressed in terms of the scattering lengths $a_{0}=\,4.468\times10^{-10}\mathrm{m}$ and $a_{1} =\,1.586\times10^{-10}\textrm{ m}$ \cite{Ivanov2001, Ivanov2002}, and 
$s^{+}m^{+}$ ($s^{-}m^{-}$) denote the spin quantum numbers of the positron(electron) pair. The s-wave scattering constraint requires that that $s^+=s^-\equiv s =0$ or $1$.  
It is convenient to express the sum in Eq. \ref{eq:U} as
\begin{eqnarray}
&&\sum_{\substack{ m^{+}, m^{-}}}U_{1} \left|1 m^{+};1 m^{-}\rangle \langle 1m^{+};1m^{-}\right|+U_0\left|00;00\rangle \langle 00;00\right|\nonumber\\
&&=U_1\mathbb{I} +(U_0-U_1)\left|00;00\rangle \langle 00;00\right|,
\end{eqnarray}
where $\mathbb{I}$ is the identity operator in the basis of $s$-wave scattering states. Next, we expand the lepton-pair ket
$|00;00\rangle=1/2\left(|\uparrow\downarrow\rangle-  |\downarrow\uparrow\rangle\right)_{e^-}\otimes\left(|\uparrow\downarrow\rangle-  |\downarrow\uparrow\rangle\right)_{e^+}$ in terms of \textit{o}-Ps and \textit{p}-Ps states. It follows that 
\begin{eqnarray}
\label{eq:psbasis}
&&\left(|\uparrow\downarrow\rangle-  |\downarrow\uparrow\rangle\right)_{e^-}\otimes\left(|\uparrow\downarrow\rangle-  |\downarrow\uparrow\rangle\right)_{e^+}\nonumber\\
&&=|\uparrow\uparrow\rangle_{{\rm{Ps}}_1}\otimes|\downarrow\downarrow\rangle_{{\rm{Ps}}_2}+|\downarrow\downarrow\rangle_{{\rm{Ps}}_1}\otimes|\uparrow\uparrow\rangle_{{\rm{Ps}}_2}\nonumber\\
&&\;-1/2\left(|\uparrow\downarrow\rangle+  |\downarrow\uparrow\rangle\right)_{{\rm{Ps}}_1}\otimes\left(|\uparrow\downarrow\rangle+  |\downarrow\uparrow\rangle\right)_{{\rm{Ps}}_2}\nonumber\\
&&\;+1/2\left(|\uparrow\downarrow\rangle- |\downarrow\uparrow\rangle\right)_{{\rm{Ps}}_1}\otimes\left(|\uparrow\downarrow\rangle- |\downarrow\uparrow\rangle\right)_{{\rm{Ps}}_2}\nonumber\\
&&=|1\rangle|-1\rangle+|-1\rangle|1\rangle-|0\rangle|0\rangle+|{\rm{p}}\rangle|{\rm{p}}\rangle\nonumber\\
&&\equiv |\Psi_{s}\rangle.
\end{eqnarray}
Substituting $|\Psi_{s}\rangle$ into Eq. \ref{eq:U}, we obtain the effective interaction represented in the basis of Ps scattering states
\begin{eqnarray}
\label{eq:U2}
U(\mathbf{r},\mathbf{r}^\prime)&&=\delta(\mathbf{r}-\mathbf{r}^\prime)\left( g_{0}\mathbb{I} +g_{1} \left|\Psi_s \rangle \langle \Psi_s \right|\right).
\end{eqnarray}
where $g_{0}=U_{1}$ and $g_{1}=(U_{0}-U_{1})/4$. 

To construct the many-body interaction Hamiltonian, we follow the standard procedure and replace the two-body interaction with a sum over field operators
\begin{eqnarray}
\mathcal{H}_{\rm int} & = & \int d^3 r d^3 r^\prime \left[ \Psi_{i}^\dagger ({\bf r}) \Psi_{j}^\dagger ({\bf r}^\prime) U_{ijkl}({\bf r}, {\bf r}^\prime)\Psi_{k} ({\bf r}^\prime) \Psi_{l} ({\bf r}) \right]. \nonumber \\
\end{eqnarray}
Performing the delta function integral, and re-arranging terms, we come to the expression in Eq. \ref{eq:H_int}.

To find the cross sections used in the semiclassical rate equations, we can use the same basis transformation procedure given above. We find the effective scattering lengths $a_{ij,kl}$ at which two Ps atoms in states $i,j$ are scattered to states $k,l$ are
\begin{eqnarray}
\label{eq:scattering length}
a_{ij,kl}=a_1 \delta_{ik} \delta_{jl}+ \frac{a_0-a_1}{4}\left\langle kl |\Psi_s \right\rangle \left\langle \Psi_s |ij \right\rangle.
\end{eqnarray}
The corresponding cross-section in the low-energy limit is then given by $\sigma_{ij,kl}=8\pi a_{ij,kl}^2$. These results are consistent with an equivalent procedure found in Ref. \cite{Ivanov2001,Ivanov2002}.

\section{Commutation relations of the $O(4)$ group}
The $O(4)$ group \cite{Judd1975} is generated by the six operators defined in Sec. \ref{sec:sym}: $S_1$, $S_2$, $S_3$ and $R_1$, $R_2$, $R_3$. These generators satisfy the required commutation relations for the $O(4)$ group 
\begin{eqnarray}
\label{eq:commutation1}
[S_a,S_b]&=&i\varepsilon_{abc}S_c, \\
\label{eq:commutation2}
 [R_a,R_b]&=&i\varepsilon_{abc}S_c, \\
\label{eq:commutation3}
 [S_a,R_b]&=&i\varepsilon_{abc}R_c.
\end{eqnarray}
An arbitrary group element is represented by
\begin{eqnarray}
\label{eq:commutation4}
D_{\mathbf{\hat{n}}_1,\mathbf{\hat{n}}_2}(\alpha_1,\alpha_2) = e^{-i\alpha_1 \hat{\mathbf{n}}_1 \cdot \mathbf{S}-i\alpha_2 \hat{\mathbf{n}}_2 \cdot \mathbf{R}},
\end{eqnarray}
where the parameters $\alpha_1, \hat{\mathbf{n}}_1$ represent the rotation angle and rotation axis for the ortho-sector, and $\alpha_2, \hat{\mathbf{n}}_2$ correspond to rotations between the ortho- and para- sectors. The group of elements spanned by $\alpha_2=0$, form a subgroup equivalent to $SO(3)$ and corresponds to physical spin rotations in the ortho- sector. Including the internal energy splitting between the ortho- and para- sectors, the symmetry of the full system is reduced from a full $O(4)$ symmetry to this $SO(3)$ subgroup.

\section{Bogoliubov-de Gennes equations}
\label{app:BdG}
We now analyze the stability of a BEC in a reference state $\psi^0$ by introducing small fluctuations $\delta \psi$:
\begin{eqnarray}
\label{eq:d_psi}
\psi_j\left(\mathbf{r},t\right)=\left(\psi_j^0\left(\mathbf{r}\right)+\delta \psi_j\left(\mathbf{r},t\right) \right)e^{-i\mu t/\hbar},
\end{eqnarray}
where $j$ is the conventional spin index used in this paper, and the fluctuation $\delta\psi_j$ has the form 
\begin{eqnarray}
\delta\psi_j\left(\mathbf{r},t\right)=u_j\left(\mathbf{r}\right) e^{-i\varepsilon t/\hbar}+v_j^*\left(\mathbf{r}\right) e^{i\varepsilon t/\hbar}.
\end{eqnarray}
When the system size is much greater than the healing length, we can assume plane wave solutions of the type $u_j\left(\mathbf{r}\right)\equiv  u_j\left(\mathbf{k}\right) e^{-i\mathbf{k}\cdot\mathbf{r}}$ and $v_j\left(\mathbf{r}\right)\equiv v_j\left(\mathbf{k}\right) e^{-i\mathbf{k}\cdot\mathbf{r}}$, and $\varepsilon\equiv\varepsilon\left(\mathbf{k}\right)$, where $\mathbf{k}$ is the characteristic wavevector of the plane wave. To calculate the fluctuations $u,v$ we substitute Eq. \ref{eq:d_psi} into the coupled GP equations (Eq. \ref{eq:GP}), linearize to first order in $\delta\psi$, and then collect terms whose phases rotate as $e^{-i\varepsilon t/\hbar}$. We obtain the Bogoliubov-de Gennes equations for $u_j$ and $v_j$
\begin{eqnarray}
\label{eq:bdg}
  H_{ji} u_i^{\left(\lambda\right)}\left(\mathbf{k}\right)+H^{'}_{ji}v_i^{\left(\lambda\right)}\left(\mathbf{k}\right) &=& \varepsilon^{\left(\lambda\right)} \left(\mathbf{k}\right)u_{j}^{\left(\lambda\right)}\left(\mathbf{k}\right) \nonumber\\
H^{'*}_{ji} u_i^{\left(\lambda\right)}\left(\mathbf{k}\right)+H^{*}_{ji}v_i^{\left(\lambda\right)} \left(\mathbf{k}\right) &=& -\varepsilon^{\left(\lambda\right)}\left(\mathbf{k}\right) v_{j}^{\left(\lambda\right)}\left(\mathbf{k}\right) \nonumber\\ 
\end{eqnarray}
where $j,\;i=1,\;0,\;-1$ and ${\rm{p}}$ are the indices for spin states, and $\lambda$ is a band index whose meaning will presently become clear. The matrices in Eq. \ref{eq:bdg} are given by 
\begin{eqnarray}
\label{eq:bdg3}
H_{ji}&=&\left(\frac{\hbar^2 k^2}{2m}-\mu\right)\delta_{ji}+g_0 \psi_j \psi_i^*(1+\delta_{ji})+\frac{\partial^2 F}{\partial \psi^*_j\partial\psi_i},\nonumber\\
H^{'}_{ji}&=&g_0 \psi_j \psi_i+\frac{\partial^2 F}{\partial \psi^*_j \partial\psi^*_i},\nonumber\\
\end{eqnarray}
where
\begin{eqnarray}
\label{eq:bdg4}
F\equiv\frac{g_1}{2}\left|2\psi_1\psi_{-1}-\psi_0^2+\psi_{\rm{p}}^2\right|^2,
\end{eqnarray}
and in Eqs. \ref{eq:bdg3}, \ref{eq:bdg4} only, we drop the superscript 0 in $\psi_j^0$. 

For each given value of $\mathbf{k}$, Eq. \ref{eq:bdg} is an eight-dimensional generalized eigenvalue problem. This is because the four spin indices are replicated in the four-dimensional matrices $H_{ji}$ and $H'_{ji}$. Thus for a given value $\mathbf{k}$, Eq. \ref{eq:bdg} has eight eigenvalues $\varepsilon^{\left(\lambda\right)}\left(\mathbf{k}\right) $ where $\lambda$ is an eight-fold index. It can be shown that the eigenvalues occur in pairs such that if $\varepsilon$ is an eigenvalue, then so is $-\varepsilon$. The eigenvalues $\varepsilon$ may in general be complex numbers, but only real eigenvalues correspond to stable fluctuations about the reference state $\psi^0$. A complex eigenvalue indicates an instability of the GP equations. At short times the amplitude of the instability grows exponentially with time constant given by $\hbar/{\rm{Im}}[\varepsilon]$. In the presence of an instability the fluctuations $\delta \psi_j$ will grow until they are no longer small. At this time, the full GP equations will be necessary to describe the evolution. The BdG equations give an accurate quantitative description of the stability of the condensate prepared in a given initial state. 

\nocite{*}
\bibliography{PsBECbib}
\end{document}